\documentclass[preprint,superscriptaddress,nofootinbib,preprintnumbers,amsmath,amssymb,notitlepage]{revtex4-1}
\usepackage[T1]{fontenc}

\usepackage[titles]{tocloft}
\cftsetindents{section}{0em}{2.5em}
\cftsetindents{subsection}{2.5em}{2.5em}

\usepackage{amsfonts}
\usepackage{mathrsfs}
\usepackage{leftidx}
\usepackage{amssymb}
\usepackage{placeins}
\usepackage{relsize}
\usepackage{slashed}

\usepackage[dvipsnames]{xcolor}
\definecolor{red}{rgb}{0.9, 0,0}
\definecolor{cerulean}{rgb}{0., 0.42,0.9}
\definecolor{navy}{rgb}{0.05, 0.05,0.8}

\usepackage[colorlinks]{hyperref}
\hypersetup{
     colorlinks   = true,
     citecolor    = red,
	linkcolor = navy
}

\usepackage{soul}
\usepackage{epsfig}
\usepackage{graphicx}               % Standard graphics package
\usepackage{url}
\usepackage{float}
\usepackage{color}

\newcommand{\be}{\begin{equation}}
\newcommand{\ee}{\end{equation}}
\newcommand{\bea}{\begin{eqnarray}}
\newcommand{\eea}{\end{eqnarray}}
\newcommand{\beq}{\begin{eqnarray}}
\newcommand{\eeq}{\end{eqnarray}}
\newcommand{\Fig}[1]{Fig.~\ref{#1}}
\newcommand{\Eq}[1]{Eq.~(\ref{#1})}

\newcommand{\Sec}[1]{Sec.~\ref{#1}}

\newcommand{\App}[1]{App.~\ref{#1}}

\newcommand{\mHe}{m_\text{He}}
\newcommand{\bfq}{{\bf q}}

\newcommand{\bfp}{{\bf p}}
\newcommand{\bfk}{{\bf k}}
\newcommand{\bfl}{{\bf l}}
\newcommand{\bfv}{{\bf v}}
\newcommand{\bfr}{{\bf r}}

\newcommand{\bfE}{{\bf E}}

\newcommand{\bfe}{\boldsymbol\epsilon}

\newcommand{\ket}[1]{| #1 \rangle}
\newcommand{\GSbra}{\langle \Psi_0 |}
\newcommand{\GSket}{ | \Psi_0 \rangle}

\newcommand{\n}{n}

\newcommand{\bigsum}{\mathlarger{\sum}}

\begin{document}

%--------------------   Previous format   ----------------------------

\title{Light Dark Matter in Superfluid Helium:\\
Detection with Multi-excitation Production}
\author{Simon Knapen}
\affiliation{Theory Group, Lawrence Berkeley National Laboratory, Berkeley, CA 94709 USA}
\affiliation{Berkeley Center for Theoretical Physics, University of California, Berkeley, CA 94709 USA}
\author{Tongyan Lin}
\affiliation{Theory Group, Lawrence Berkeley National Laboratory, Berkeley, CA 94709 USA}
\affiliation{Berkeley Center for Theoretical Physics, University of California, Berkeley, CA 94709 USA}
\author{Kathryn M. Zurek}
\affiliation{Theory Group, Lawrence Berkeley National Laboratory, Berkeley, CA 94709 USA}
\affiliation{Berkeley Center for Theoretical Physics, University of California, Berkeley, CA 94709 USA}

\begin{abstract}
We examine in depth a recent proposal to utilize superfluid helium for direct detection of sub-MeV mass dark matter.
For sub-keV recoil energies, nuclear scattering events in liquid helium primarily deposit energy into long-lived phonon and roton quasiparticle excitations.
If the energy thresholds of the detector can be reduced to the meV scale, then dark matter as light as $\sim$ MeV can be reached with ordinary nuclear recoils. If, on the other hand, two or more quasiparticle excitations are directly produced in the dark matter interaction, the kinematics of the scattering allows sensitivity to dark matter as light as $\sim$ keV at the same energy resolution. We present in detail the theoretical framework for describing excitations in superfluid helium, using it to calculate the rate for the leading dark matter scattering interaction, where an off-shell phonon splits into two or more higher-momentum excitations.  We validate our analytic results against the measured and simulated dynamic response of superfluid helium.   Finally, we apply this formalism to the case of a kinetically mixed hidden photon in the superfluid, both with and without an external electric field to catalyze the processes.
\end{abstract}

\date\today

%\end{abstract}
%
%%\preprint{}
%
%
\maketitle
\flushbottom
%
%\hrulefill
\clearpage
\tableofcontents
%--------------   End of previous format  -------------------------

%%%%%%%%%%%%%%%%%%%%%
\section{Introduction}
%%%%%%%%%%%%%%%%%%%%%

Weakly Interacting Massive Particles (WIMPs) with a mass of ${\cal O}(100)$ GeV have been one of the leading dark matter (DM) candidates for the past few decades.  However, recent null results in direct detection and collider experiments now provide strong motivation to extend the scope of our models and searches as much as possible. 
In addition, theoretical advances have shown that there are a variety of models for sub-GeV dark matter that are only now beginning to be explored. Such dark matter may reside in a low mass hidden sector (or ``hidden valley'') at the MeV-GeV scale \cite{Strassler:2006im}, with either strongly or weakly interacting dynamics~\cite{Boehm:2003hm,Pospelov:2007mp,Hooper:2008im,Feng:2008ya}.  These particles can be invisible to production at colliders, but give rise to large scattering cross-sections in direct detection experiments.  They are moreover well-motivated in Asymmetric Dark Matter~({\it e.g.}~\cite{Kaplan:2009ag}), supersymmetric hidden sectors~\cite{Morrissey:2009ur,Baumgart:2009tn}, and SIMP dark matter~\cite{Hochberg:2014dra}, to name a few.

The theoretical progress in identifying sub-GeV dark matter has been accompanied by effort to experimentally probe such light dark matter~\cite{Alexander:2016aln}.
Among the various ways to detect dark matter, existing direct detection experiments have traditionally focused on nuclear recoils from WIMPs, with typical recoil energies of 10-100 keV. Rapid progress in recent years has thus produced strong limits on DM-nucleon scattering in the 10-100 GeV mass range~\cite{Aprile:2012nq,Akerib:2015rjg,Agnese:2014aze,Tan:2016zwf}, tightly constraining many well-motivated models of WIMP dark matter. To improve sensitivity to lower mass DM, a number of these experiments have successfully developed techniques that lower the nuclear recoil thresholds below $\sim$ keV. This has been implemented for example in CDMSlite~\cite{Agnese:2015nto} and CRESST~\cite{Angloher:2015ewa}, which are sensitive to GeV-scale dark matter.

For a given deposited energy, sensitivity to lighter dark matter can be obtained by scattering from electrons, rather than nuclei.  This is because in elastic scattering with the target at rest, the deposited energy is $\omega = { q}^2/2 m_T$, where $m_T$ is the target mass and the momentum transfer $q \sim \mu_r v_X$ is given by the dark matter velocity $v_X$ and the dark matter-target reduced mass $\mu_r$.  
The first effort in this direction utilized an electron ionization process in XENON10, deriving a constraint on electron interaction cross-sections for DM heavier than 10 MeV~\cite{Essig:2012yx}.  For this mass, the DM possesses the minimum kinetic energy needed to ionize an electron from xenon, $\sim$ 12 eV. In the future, SuperCDMS may have sensitivity to MeV-scale DM, on account of the smaller $\sim$ eV excitation energy set by the band gap of the semiconductor~\cite{Essig:2011nj,Lee:2015qva,Essig:2015cda}.   (SuperCDMS may also probe unexplored parameter space for light bosonic DM with eV-keV mass through an absorption process~\cite{Hochberg:2016sqx,Bloch:2016sjj}.)
Other small gap materials may also make good targets for MeV-GeV mass dark matter in scattering, most notably graphene~\cite{Hochberg:2016ntt}, giving access to directional information, and crystal scintillators~\cite{Derenzo:2016fse}. 

To reach DM lighter than an MeV, new ideas are needed.   The first proposal sensitive to keV-scale DM considered superconductors \cite{Hochberg:2015pha,Hochberg:2015fth} for DM-electron scattering. A conventional superconductor has a small $\sim$0.3 meV electron band gap and a large electron Fermi velocity, $v_F \sim 10^{-2} c$; these two facts combined kinematically allow access to keV mass DM (carrying a meV of kinetic energy).  It was also shown that these targets have a remarkable sensitivity to bosonic DM in the meV-eV mass range via absorption on electrons, followed by phonon emission~\cite{Hochberg:2016ajh}. Aside from the $\sim$meV electron band gap, superconductors have another property which can allow for detection of small energy depositions; a DM scattering that breaks a Cooper pair will give rise to long-lived quasiparticle excitations (which behave very much like an electron). In a very clean superconductor, excitations created in the bulk can then be detected in sensors at the surface of the target.  
Among the experimental challenges to implementing this idea, it is necessary that the energy resolution be improved significantly, down to the meV scale.

In this paper, we turn to a new proposal to detect keV-MeV scale DM via nuclear recoils in superfluid helium, first discussed in Ref.~\cite{Schutz:2016tid}. Similar to the superconductor target, the low-energy degrees of freedom in superfluid helium are long-lived quasiparticles. These quasiparticle excitations (called phonons, rotons and maxons) are collective modes in the fluid, analogous to sound waves in the long-wavelength limit.  These modes are produced by nuclear scattering, and have been extensively probed by neutron scattering experiments on superfluid helium. Since a large fraction of the deposited energy in a low-energy nuclear scattering is converted to phonons and rotons, it may then be possible to detect dark matter as light as MeV via regular nuclear recoils if experimental thresholds can be lowered to $\sim 10$ meV.  This is because MeV mass dark matter deposits $\sim$ meV of energy in a nuclear recoil process, but this energy is amplified by $\sim 10$ meV through the evaporation of the excitation at the surface of the superfluid.  For a discussion of experimental aspects of a liquid helium detector, and possibilities for detecting the phonons and rotons, see for example Refs.~\cite{Guo:2013dt,mckinseyslac}.

\begin{figure}
\includegraphics[width=0.4\textwidth]{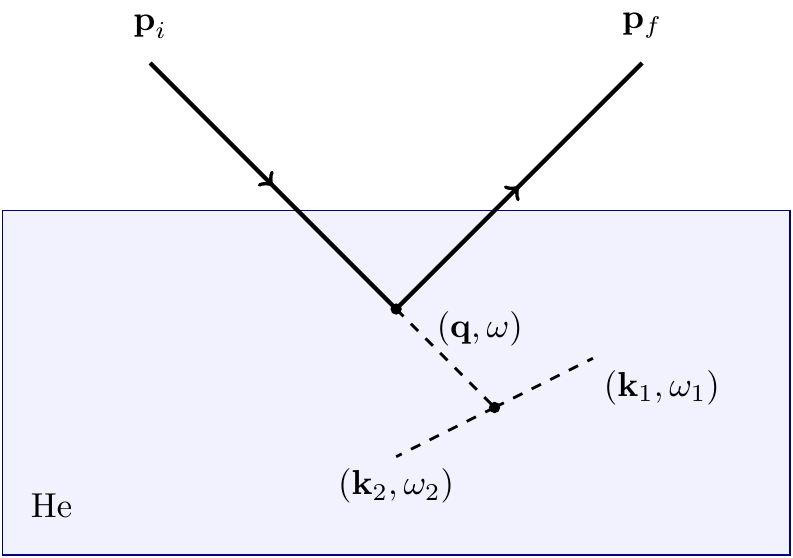}\hfill
\includegraphics[width=0.4\textwidth]{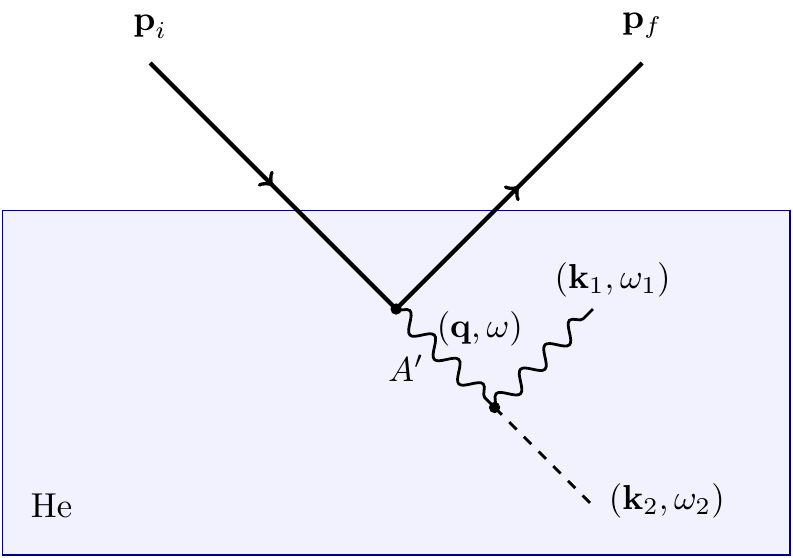}
\caption{Leading order contribution to DM scattering via quasiparticle production in superfluid helium.  {\bf(left)}\ \ For scattering through a contact interaction, the off-shell intermediate excitation splits into two, nearly back-to-back on-shell excitations.  {\bf (right)} \  For scattering through an intermediate hidden photon $A'$, the hidden photon splits into a real photon, which carries nearly all the energy, and a fluid excitation, which carries nearly all the momentum. \label{fig:scattering}}
\end{figure}

The idea of Ref.~\cite{Schutz:2016tid} was to probe lower mass DM, in the keV-MeV range, by taking advantage of  multi-excitation production in superfluid helium. For these low masses, which have correspondingly small momentum $\lesssim$ keV, the DM couples directly to the collective quasiparticle modes. However, the kinematics prohibit the creation of a single excitation with energy above a meV. The underlying reason is that the dark matter velocity is much larger than the typical sound speed in the fluid, such that the typical energy and momentum transfer for sub-MeV DM cannot match the dispersion relation of a single, on-shell excitation. However, by considering the  process of emitting two or more excitations, it is possible to deposit energies larger than $\sim$ meV even with the small momentum transfers characteristic of such light dark matter. The final state excitations are higher-momentum excitations and very nearly back-to-back. The left panel of \Fig{fig:scattering} illustrates this process. 

The purpose of this paper is two-fold.  First, we amplify the discussion of Ref.~\cite{Schutz:2016tid}, providing many more details of the theory utilized for computing the multi-excitation scattering rate.  We update the analytic calculation of Ref.~\cite{Schutz:2016tid} with the measured structure factor for the leading-order scattering rate, and again compare against the available computations of the literature.  While neutron scattering data and detailed numerical simulations have been studied in some parts of the multi-excitation phase space, the fluid response for DM with mass below $\sim 100$ keV  rests partially in previously unconsidered regimes of momentum transfer and energy deposition. This therefore requires some theoretical understanding of the rates. 
Second, we elaborate on the reach for a simplified model of dark matter coupling to nuclei via a new mediator and compare with existing constraints. We additionally consider scattering and absorption via hidden photons, where the final state is a real photon plus a fluid excitation  (see right panel of \Fig{fig:scattering}). In this case, the real photon carries away the bulk of the energy, while the fluid excitation absorbs most of the momentum. However, since the net electric charge of a helium atom is screened at the wavelengths of interest, we find that the reach for this case is not competitive with existing stellar constraints.   

We introduce the basic elements of the theory for superfluid helium in \Sec{sec:SuperfluidHelium}, beginning with a broad introduction to the nature of quasiparticle excitations in the superfluid. In order to calculate the two-excitation process, we employ the correlated basis function formalism, standard in the liquid helium literature, and derive the three-excitation matrix element. 
\App{app:secondquant} provides an alternative formulation in terms of second quantization, and \App{app:overlap} fills in extra details of calculating the three-excitation matrix element.  In \Sec{sec:DMscattering}, we turn to a comparison of numerical calculations of the multi-excitation process, applying our results to derive the sensitivity of a liquid helium target to light DM. The results here focus on DM scattering via a mediator that couples to the nucleus. In \Sec{sec:DarkPhoton}, we discuss scattering and absorption processes involving a hidden photon, which couples to liquid helium via its polarizability. 
We conclude in \Sec{sec:Conclusions}.

%%%%%%%%%%%%%%%%%%%%%%%%%%%%%%%%%
\section{Theory of superfluid helium}\label{sec:SuperfluidHelium}
%%%%%%%%%%%%%%%%%%%%%%%%%%%%%%%%%

A unique property of helium in the superfluid phase is the nature of the elementary excitations. At long wavelengths ($1/\lambda \lesssim \textrm{few keV}$), the elementary degrees of freedom are no longer single-atom excitations. Instead, the elementary excitations are acoustic phonon modes, a collective mode which is equivalent to a density perturbation at long wavelengths. 
The quasiparticle nature of the phonon modes is also essential for dark matter detection. While phonon modes are present even at temperatures above $T_c = $ 2.17 K, the critical temperature for the superfluid phase transition, it is only well below $T_c$ that the width of the phonon mode becomes narrow. In this regime these phonon modes are the only excitations present and they can be thought of as nearly stable quasiparticles. Since a large fraction of the energy deposited in a low-energy dark matter scattering will be in the form of phonons, it is important that these excitations be long-lived states that can propagate to the surface of the liquid and be measured, for a $\sim 10 \ $cm$^3$ volume (or 1 kg) of liquid helium (see Refs.~\cite{Maris,WoodsCowley1972} for a discussion on phonon lifetimes).

In this section, we describe the theory for superfluid helium needed to calculate the production of multiple excitations in the liquid. Due to the strongly-interacting nature of the liquid, the underlying microscopic theory for superfluid helium is not completely understood. However, somewhat phenomenological methods have been proposed which can successfully reproduce many features of the data. The basic idea behind these methods goes back to Feynman in 1954~\cite{Feynman54}, and starts with a posited form for the ground state $| \Psi_0 \rangle$, or equivalently a wavefunction for an $N$-atom system. While determining the form of the ground state is difficult (though it can be tested by comparison with data), excited states are momentum eigenstates that are written simply as the number density operator acting on the ground state, $|\bfq \rangle \propto n_{\bfq} |\Psi_0\rangle$. This starting point will then allow us to calculate the creation of excitations of the liquid, even without complete knowledge of the full ground state.  We will compare this approach with more complete calculations available in the literature. For a thorough discussion of the various theoretical descriptions of excitations in liquid helium, see also Refs.~\cite{Griffin,glyde1994excitations,Feenberg}.

%%%%%%%%%%%%%%%%%%%%%%%%%%%%%%%%%%%%%%%%%%%%%%%%
\subsection{Bijl-Feynman relation for single-excitations \label{sec:BijlFeynman}  }
%%%%%%%%%%%%%%%%%%%%%%%%%%%%%%%%%%%%%%%%%%%%%%%%

Much of our knowledge of the excitations in a strongly-interacting quantum fluid (such as superfluid helium) comes from the \emph{dynamic structure function} $S({\bfq},\omega)$, which describes the response of the liquid to a density perturbation with momentum transfer ${\bfq}$ and energy deposited $\omega$.  For instance, $S(\bfq,\omega)$ can be directly measured in neutron scattering by measuring the differential cross section:
\beq\label{eqn:neutronscattering}
\frac{d^2 \sigma}{d\Omega d \omega} = b_n^2 \frac{p_f}{p_i} S({\bfq},\omega),
\eeq
where ${\bf p}_i$ and ${\bf p}_f$ are the initial and final momenta of the scattered neutron, $\bfq = \bfp_f - \bfp_i$, and $b_n$ is the  scattering length of a neutron on an individual helium nucleus. 

The dynamic structure function thus depends on the matrix element for the creation of a quasiparticle with momentum $\bfq$ and energy $\omega$. Concretely, $S(\bfq,\omega)$ is defined as: 
\beq\label{eq:dynamicstructure}
	S(\bfq,\omega) \equiv \frac{1}{ n_0} \sum_\beta |\langle \Psi_\beta | \n_{\bfq} \GSket |^2 \delta(\omega - \omega_\beta) ,
	\label{eq:Sqw}
\eeq 
where the final states in the scattering are denoted as $|\Psi_\beta \rangle$ with energy $E_\beta$, the ground state $ \ket{ \Psi_0}$  has energy $E_0$, and $\omega_\beta = E_\beta - E_0$. Here $\n_{\bf q}$ is the Fourier transform of the density operator (in real space, \mbox{$n(\bfr)=\sum_i \delta^{(3)}(\bfr - \bfr_i)$}),
\beq
	\label{eq:densitydef}
	\n_{\bfq}\equiv \frac{1}{\sqrt{V}} \sum_{i=1}^N\exp(i{\bfq}\cdot {\bf r}_i),
\eeq
and ${\bf r}_i$ are the coordinates of the individual helium atoms in the fluid.
We take an arbitrary quantization volume $V$, with $N$ the number of He atoms in the volume; physical results will only depend on the average number density $n_0 = N/V$. To facilitate some of the later computations, we will occasionally go to the continuum limit by replacing
$
\sum_{\bfq}\rightarrow V/(2\pi)^3\, \int d^3\bfq$ and $\delta_{\bfq,\bfq'}\rightarrow(2\pi)^3/V\,\delta^{(3)}(\bfq-\bfq').
$

The reason neutron scattering (or dark matter scattering) couples to density fluctuations can be understood by considering the potential $V({\bf r})$ seen by a neutron in the liquid, 
\begin{align}
	V(\bfr) = \frac{2\pi b_n}{m_n} \sum_i \delta^{(3)}( \bfr - \bfr_i) = \frac{2\pi b_n}{m_n}  n(\bfr),
	\label{eq:potential}
\end{align}
assuming a hard-sphere interaction and neutron mass $m_n$. This is the underlying justification for \Eq{eqn:neutronscattering}, which we derive in \Sec{sec:DMsetup} for the case  of DM scattering. In particular, we similarly obtain such a potential for dark matter by coupling the DM to helium atoms, with $b_n/m_n \to b_X/m_X$, with $b_X$ and $m_X$ the dark matter mass and scattering length, respectively. 

The dynamic structure function $S(\bfq,\omega)$ is thus crucial to understand the response of superfluid helium to dark matter scattering. While it can be obtained from neutron scattering data at moderate momentum transfer ($q \gtrsim 0.1 $/\AA, corresponding to $q \sim 0.2$ keV in units where $q = 2 \pi /L$), a certain level of theoretical control is also possible.  As we will see, this theoretical control will be crucial for extrapolating the dynamic structure function to lower momentum transfers, which is necessary to compute the scattering rate when the DM is lighter than $\sim 100$ keV. 

\begin{figure}[bt]\centering
\includegraphics[width=0.47\textwidth]{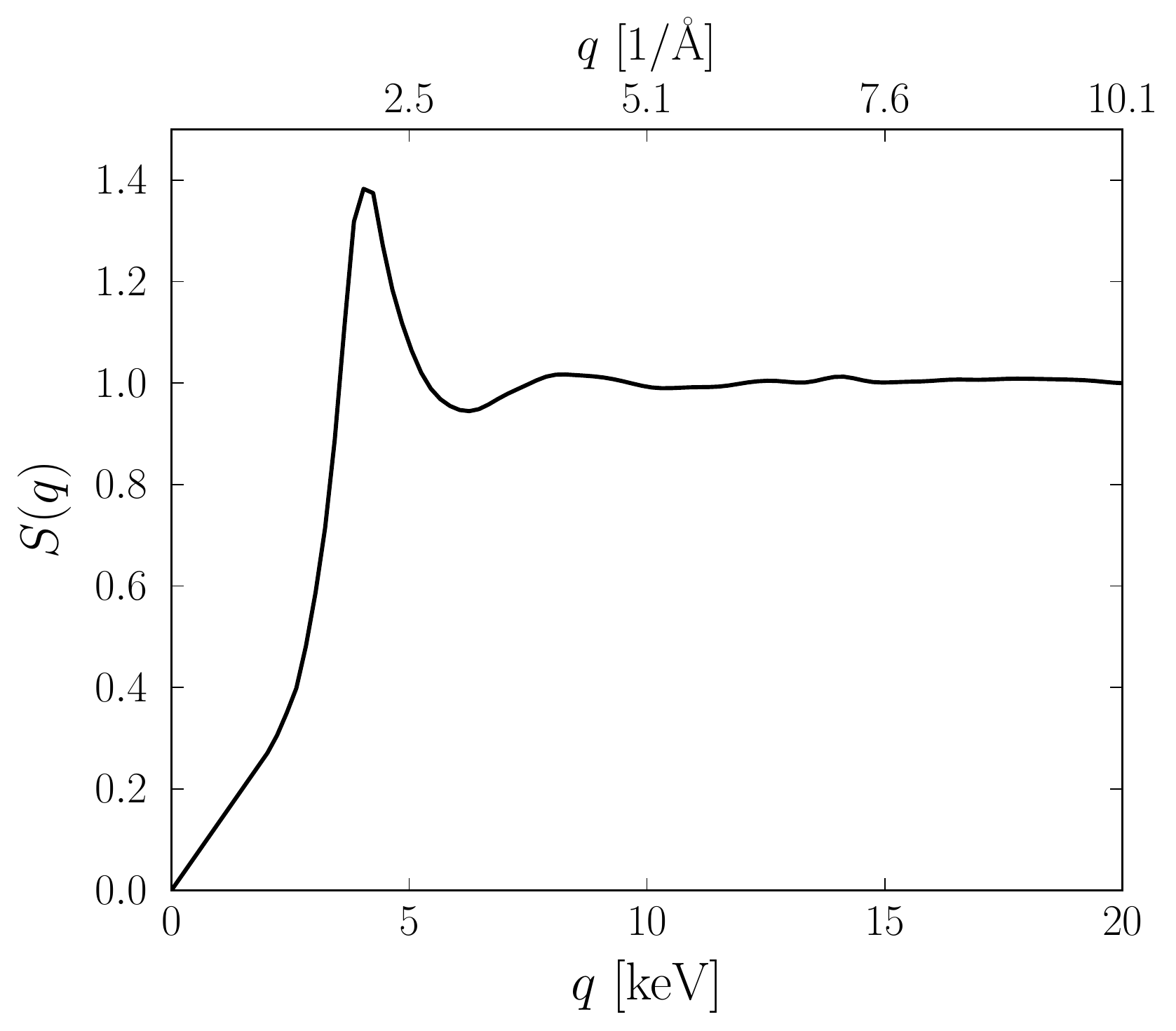}
\hspace{0.25cm}
\includegraphics[width=0.5\textwidth]{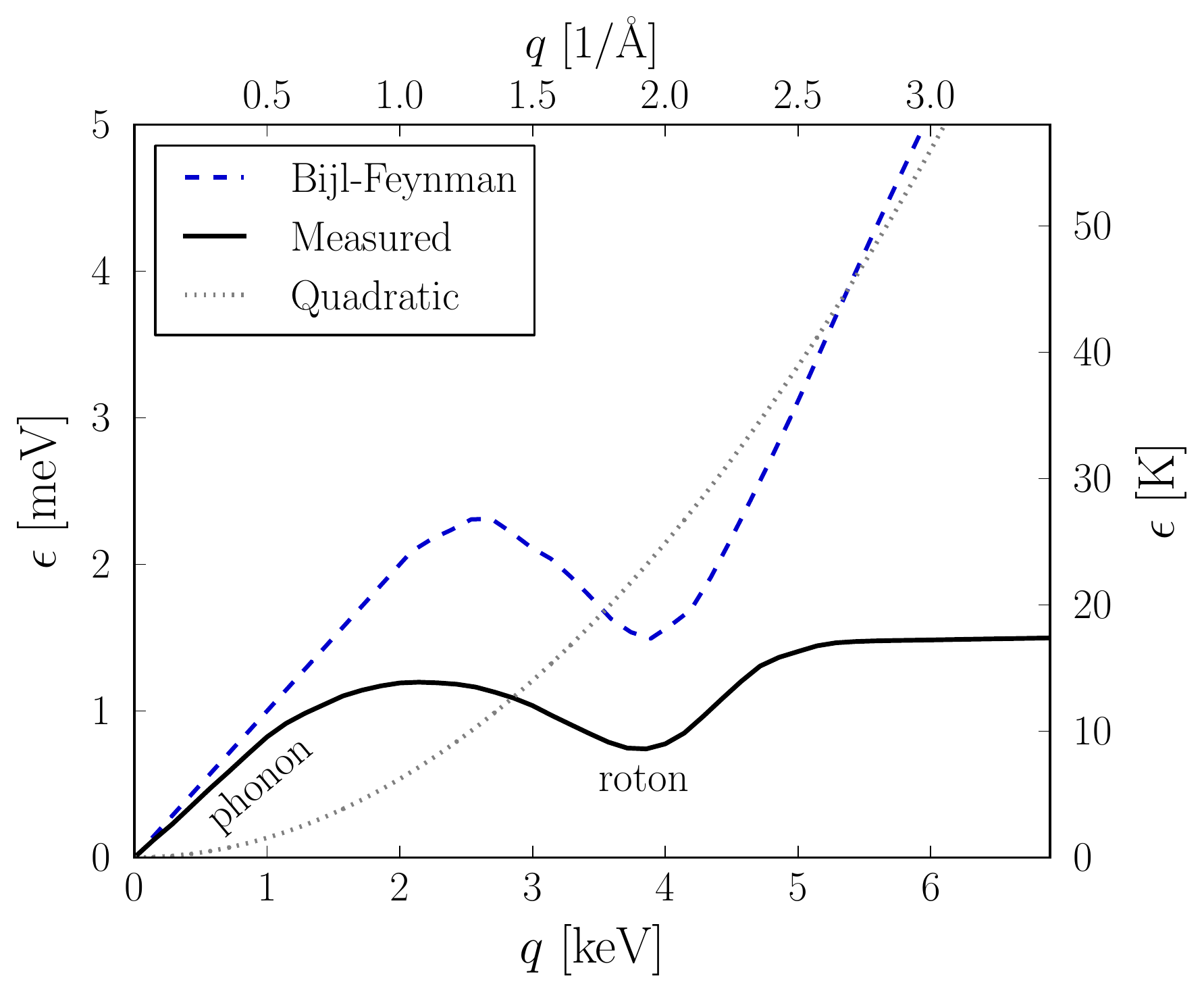}
\caption{
	{\bf (left)} Interpolation of the data for the static structure function $S(\bfq)$ (at $T= 1$ K) from neutron scattering experiment~\cite{Svensson:1980zz}. In the small $\bfq$ limit, $S(\bfq)$ behaves linearly according to \Eq{eq:Sklinear}. 
   {\bf (right)} We compare the measured dispersion curve for single excitations in superfluid helium (solid black line) with the Bijl-Feynman relation for excitations, $\bfq^2/(2 \mHe S(\bfq))$ (dashed blue line). The measured dispersion curve~\cite{Krotscheck2015} comprises the phonon modes at low $\bfq$ and the maxon and roton at high $\bfq$ (in particular, the modes at around $\bfq \sim4$ keV where $\epsilon(\bfq)$ reaches a local minimum is called the roton), but does not include the broad multi-excitation response centered around the free-particle dispersion at high $\bfq$.  In the Bijl-Feynman theory, which does track the quadratic dispersion at high $\bfq$ (shown as the dotted black line), these high $\bfq$ modes are treated as single-particle excitations.   
 \label{fig:dispersion}  }
\end{figure}

The leading order contribution to $S(\bfq,\omega)$ is given by the probability to create a single on-shell quasiparticle excitation. One of the earliest theories of the single excitation spectrum, due to Bijl~\cite{Bijl1940} and Feynman~\cite{Feynman54}, applies the variational method to understand the shape of the dispersion curve. Concretely, the trial wavefunction for a single excitation is given by
 \beq
 |\bfq\rangle = \frac{1}{ \sqrt{\n_0 S({\bfq})} }  \n_{\bfq} \GSket,
 \label{eq:singleexcitation}
 \eeq
 with $n_{\bf q}$ defined in \Eq{eq:densitydef}, and
where the \emph{static structure function}  $S({\bfq})$ is defined by
\begin{align}\label{eq:Sqdefinition}
	S({\bfq})\equiv\frac{1}{\n_0} \GSbra \n_{-\bfq} \n_{\bfq}  \GSket.
\end{align} 
where $S({\bfq})$ is a function only of  $q = | \bfq|$, and its appearance in the definition of the state ensures that $\langle \bfq|\bfq\rangle=1$. The left panel of \Fig{fig:dispersion}  shows the experimentally measured $S({\bfq})$ in helium, which is linear in $|\bfq|$ at small momentum and approaches 1 at high momentum. In the limit of a single excitation, which does not split to multi-excitations, $S({\bfq})$ is related to the dynamic structure function by
\bea
S(\bfq,\omega) & \approx & \frac{1}{n_0}\big| \langle\bfq| \n_{\bfq} \GSket \big|^2 \delta(\omega - \epsilon_0({\bfq})) \\ \nonumber
& = & S({\bfq}) \delta(\omega - \epsilon_0({\bfq})),
\eea
where $\epsilon_0(\bfq)$ is the energy of $|\bfq \rangle$, which we will refer to as the Bijl-Feynman energy.

Since the state in \Eq{eq:singleexcitation} is by construction orthogonal to the ground state, the variational method dictates that  its energy $\epsilon_0({\bfq})$ provides an upper bound on the true energy eigenvalue $\epsilon({\bfq})$.   As we will shown in the next section, the single excitation energy is
\bea\label{eq:bijlfeynman}
\epsilon_0({\bfq})\equiv \; \langle \bfq|H - E_0|\bfq\rangle=\frac{\bfq^2}{2\mHe S({\bfq})} \geq \epsilon ({\bfq}),
\eea
with $H$ the Hamiltonian and $E_0$ the ground-state energy. The factor $S(q)$ comes from the normalization of the states in \Eq{eq:singleexcitation}, and in the limit of a free Bose gas $S(\bfq)\rightarrow 1$.
The Bijl-Feynman theory for $\epsilon_0({\bfq})$ produces the single resonance curve shown in the right panel of \Fig{fig:dispersion}, and approaches the free-particle quadratic dispersion at high $q$.  For comparison, we also show the measured dispersion curve for single-resonance excitations in \Fig{fig:dispersion}. We see that the Bijl-Feynman energy agrees roughly with the measured energy at long wavelengths, where the excitations can be identified with sound waves (phonons) with energy $\epsilon_0(\bfq)=c_s |\bfq|$, where $c_s\approx2.4 \times 10^4\;\mathrm{cm}/\mathrm{s}$ is the sound speed. 
In this regime, the static structure factor is then linear in the momentum with 
\begin{equation}\label{eq:Sklinear}
	S(\bfq)\approx\frac{|\bfq|}{2\mHe c_s}\,.
\end{equation}
However, as the curve reaches a maximum and begins to turn over (the maxon and roton regions), the agreement no longer persists, and is not even qualitatively correct as the dispersion curve reaches a plateau.

%%%%%%%%%%%%%
%%%%%%%%%%%%%

The original Bijl-Feynman theory contains, however, no multiphonon response.  More generally, the dynamic structure function will contain both the single pole, with strength $Z(\bfq)$, and a continuum component, $S_m$:
\beq
S(\bfq,\omega) = Z(\bfq)\delta(\omega - \epsilon(\bfq)) + S_m(\bfq,\omega), 
\label{eq:MultiResponse}
\eeq
and the static structure function now satisfies the more general relation $S({\bfq}) = \int d\omega\, S(\bfq,\omega)$. Any  large deviation of $Z(\bfq)$ from $S(\bfq)$ indicates that the state defined in \Eq{eq:singleexcitation} is no longer a good approximation to the single-excitation state, which will be the case in the roton region.
The continuum component $S_m$ results from multi-excitation production in the medium.   These multi-excitation modes are also important for computing the correct single resonance dispersion curve through radiative corrections to the propagator. It is the multi-excitation response  $S_m(\bfq,\omega)$ that we will focus on in the rest of this section.

%%%%%%%%%%%%%%%%%%%%%%%%%%%%
\subsection{Hamiltonian formulation\label{sec:freeham}}
%%%%%%%%%%%%%%%%%%%%%%%%%%%%

We now lay out the ingredients to describe phonon interactions, focusing on the elements needed  to compute $S_m(\bfq,\omega)$. We follow the \emph{correlated basis function} formalism, which we briefly review here. This formalism adopts the Bijl-Feynman approach, positing that particle correlations are primarily contained in the ground state wavefunction. Given the exact ground state, excited states are obtained simply with repeated applications of the density operator. Following Ref.~\cite{JacksonFeenberg}, we define a lowest-order set of basis states using the Bijl-Feynman states:
\begin{align}
\label{eqn:singleparticle}
|\bfq\rangle^0&\equiv \frac{1}{\sqrt{n_0 S(\bfq)}}\, \n_\bfq | \Psi_0 \rangle\\
\label{eqn:twoex}
|\bfq_1,\bfq_2\rangle^0 & \equiv \frac{1}{\sqrt{n_0 S(\bfq_1)}}\frac{1}{\sqrt{n_0 S(\bfq_2)}}\, \n_{\bfq_1} \n_{\bfq_2} \ket{\Psi_0} .
\end{align}
$\ket{\Psi_0}$ is full ground state of the interacting system.
Importantly, the states here are not orthogonal and hence phonon number is not conserved.  Instead, the propagating excitations are superpositions of these states. We deal with this complication in the following section.

To compute the energies and matrix elements, we require an interaction Hamiltonian. This Hamiltonian may either be written as the effective theory of a quantum fluid, or in terms of the microscopic degrees of freedom.  Let us first consider the fluid Hamiltonian, which directly allows for a second-quantized approach, see {\it e.g.}~\cite{stephen1969raman}.  (This discussion most closely follows Ref.~\cite{Schutz:2016tid}.) Here we elevate the status of the density fluctuation $n_{\bfq}$ to independent operators which create excitations in the fluid, and consider an effective Hamiltonian for these fluid degrees of freedom,
\bea\label{eq:hydroham}
H&=&\int d^3 \bfr\,\left(\frac{1}{2}\mHe\, \bfv \cdot \n {\bfv}+ \mathcal{V}(\n) \right).
\eea
By expanding in the density and velocity fluctuations, the system can be approximated to leading order as a harmonic oscillator with Hamiltonian
\beq\label{eq:hamfree}
H_0  &=&  \frac{1}{2} \sum_{\bfq}  \, \mHe\, \n_0 \bfv_\bfq\cdot \bfv_{-\bfq} + \phi(\bfq) \n_{\bfq} \n_{-\bfq}  \, ,
\eea
where $\phi(\bfq)\equiv \delta^2 \mathcal{V}/\delta\n_\bfq^2$ can be thought of as a momentum dependent force constant.  As we show in \App{app:secondquant}, this Hamiltonian lends itself to canonical quantization of the $n_{\bfq}, {\bf v}_{\bfq}$ variables, and \Eq{eq:hamfree} be can expressed in terms of creation and annihilation operators
\beq\label{eq:H0creationbody}
H_0 = \sum_\bfq \epsilon_0(\bfq)\left(a_{\bfq}^\dagger a_{\bfq} + \frac{1}{2}\right).
\eeq
The single-excitation energy is simply $ ^0\langle\bfq|H_0 - E_0 |\bfq \rangle^0  =\epsilon_0(\bfq)$.
The three-excitation interaction vertex and corrected energy eigenvalues can then be obtained by expanding \Eq{eq:hydroham} to higher order in the density and velocity fluctuations.
While this setup may be more familiar to a particle physicist, it is less convenient for our purposes. In particular, the ground state in the fluid is nontrivial: in a medium, quantum fluctuations require us to consider an active vacuum, where the asymptotic states of the strongly-interacting fluid are not well-approximated by the free states of a weakly-interacting system. This effect can be accounted for in the second-quantized quantum fluid formalism by correcting the ground state order by order, as we show in \App{app:secondquant}, although the calculation is somewhat cumbersome. 
 
In practice, matrix elements are often derived more simply in a first-quantized formulation of the microscopic theory, which has the advantage, as we will see, that knowledge of the ground state is not required to compute the matrix element that we are interested in. Given the energies and vertices computed in this approach, one can of course construct an equivalent second-quantized, quantum fluid Hamiltonian, which may be more convenient for certain scattering and self-energy calculations. 
The first-quantized microscopic Hamiltonian is given by
\beq\label{eq:hamfeenberg}
H=\sum_{i} \left(-\frac{\nabla_i^2}{2\mHe} \right) + \mathcal{V}(\{\bfr_i\}),
\eeq
where the sum runs over all $N$ particles in the fluid. Writing $\psi_0( \{ {\bfr_i} \})$ as the wavefunction corresponding to the ground state $|\Psi_0\rangle$, we require that  $ H \psi_0 = E_0 \psi_0 $ such that $\psi_0$ is the exact ground state of the full Hamiltonian. For a translationally-invariant system, we thus find that the ground state energy is $ E_0 = \langle \Psi_0 | \mathcal{V}(\{\bfr_i\}) \ket{\Psi_0} $. 

We can show that this formulation also gives the Bijl-Feynman energy in \Eq{eq:bijlfeynman}.  Using $\langle ... \rangle  \to \int d^3 \bfr_1\, ...\bfr_N$ with integration over the coordinates of all atoms, and acting with $H - E_0$ on the wavefunction $|\bfq\rangle^0 \rightarrow n_{\bfq} \psi_0/\sqrt{\n_0 S(\bfq)}=\sum_i e^{i \bfr_i \cdot\bfq}\psi_0/\sqrt{N\,S(\bfq)}$, 
\begin{align}
	\leftidx{^0}\langle \bfq| H - E_0  | \bfq \rangle^0 &= \frac{1}{N S(\bfq)}  \sum_{i,\ell} \int d^3 \bfr_1\, ...\bfr_N\ \psi_0 e^{-i \bfq\cdot \bfr_i}   \left[ H - E_0 \right] \left( e^{i \bfq\cdot \bfr_\ell} \psi_0\right)  \nonumber \\
  &=\frac{1}{2\mHe} \frac{1}{N S(\bfq)}  \sum_{i,j,\ell} \int d^3 \bfr_1\, ...\bfr_N\ \psi_0 e^{-i \bfq\cdot \bfr_i}   \left[ -   \psi_0 \left(\nabla_j^2 e^{i \bfq\cdot \bfr_\ell} \right)-  2  \left(\nabla_j \psi_0 \right) \left(\nabla_j e^{i \bfq\cdot \bfr_\ell}\right)  \right] \nonumber  \\
	& = \frac{1}{2\mHe}\frac{1}{N S(\bfq)}\sum_{i,j,\ell}  \int d^3 \bfr_1\, ...\bfr_N\  \psi_0^2 \left(  \nabla_j e^{-i \bfq\cdot \bfr_i} \right) \left( \nabla_j e^{i \bfq\cdot \bfr_\ell} \right) =\frac{q^2}{2\mHe S(\bfq)}
\end{align}
where we used  $ H \psi_0 = E_0 \psi_0 $ and rearranged the derivatives with partial integration.  We have also assumed $\psi_0$ is a properly normalized, real wavefunction, $\int d^3 \bfr_1\, ...\bfr_N (\psi_0)^2 = 1$. (Notice that the dependence on the unknown forms of $\psi_0$ and $ \mathcal{V}(\{\bfr_i\})$ dropped out.) While we have only computed the average energy for a given state, it can furthermore be shown that $| \bfq \rangle$ approaches an exact eigenstate of $H$ in the $q \to 0$ limit~\cite{DavisonFeenberg}.

%%%%%%%%%%%%%%%%%%%%%%%%%%%%%%%%%%
\subsection{Three-excitation vertex \label{sec:threephononvertex}  }
%%%%%%%%%%%%%%%%%%%%%%%%%%%%%%%%%%

In the previous section, we defined a single excitation state which we regard as an approximately free quasiparticle, as well as multi-excitation states which are products of the single excitations. However, an important subtlety in treating a non-dilute, strongly interacting fluid is that the asymptotic states do not have a well defined particle number. In particular, the states defined so far are not orthogonal, and we must   first define an orthogonal basis of states before considering the three-excitation vertex. In other words, to correctly calculate the cross section, we need to compute the matrix element for states which are long-lived compared to the time-scale set by the interaction Hamiltonian. 
 This way the factorization principle allows us to compute the total rate without detailed knowledge of the ultimate fate of the external states in the matrix element.\footnote{This is a familiar concept in hard parton scattering in QCD, where we can compute the leading order, total inclusive cross section without detailed knowledge about the shower and hadronization.} 

A set of orthogonal states can be obtained by performing a Gram-Schmidt rotation to orthogonalize the basis. Concretely, we start with the same single excitation state  $|\bfq\rangle^0 = \n_{\bfq} \ket{\Psi_0} /\sqrt{n_0 S(\bfq)}$ and then define a set of orthogonal states relative to $|\bfq\rangle^0$ by
\begin{align}
| \bfq\rangle &\equiv| \bfq\rangle^0 \\
 |\bfq_1,\bfq_2 \rangle &\equiv \ket{\bfq_1,\bfq_2}^0-\sum_{\bfq'} \langle \bfq' | \bfq_1,\bfq_2 \rangle^0\, |\bfq' \rangle.\label{eq:gramschmidt2}
\end{align}
In what follows we alway drop the $^0$ superscript for the single particle state, since it is by construction identical to the corresponding state in the orthogonalized basis.
We identify this new basis of states with the orthogonal eigenstates of the quadratic Hamiltonian for the quasiparticles, which will be corrected by the cubic interactions derived below.

The unknown particle correlations of the strongly coupled fluid are now conveniently packaged in the $\langle \bfq'| \bfq_1,\bfq_2 \rangle ^0$ matrix element, which we will discuss later in this section.  In the microscopic Hamiltonian of \Eq{eq:hamfeenberg}, this overlap term encodes the unknown potential term which dictates the correlations of particles in the ground state. For the quantum fluid effective Hamiltonian in \Eq{eq:hydroham}, the same information is encoded in the interactions coming from both the kinetic term, the unknown potential and possible matching terms encoding the unknown short distance physics. (In this sense one may roughly think of the overlap term  $\langle \bfq'| \bfq_1,\bfq_2 \rangle ^0$ as a counterterm which enforces the orthogonality of the renormalized states.)

To compute the three-excitation matrix element, we again use $\delta H = H - E_0$ with the Hamiltonian given in \Eq{eq:hamfeenberg} and with $E_0$ the ground state energy:
\bea
	\langle \bfq - \bfk , \bfk | \delta H | \bfq \rangle = \  \leftidx{^0}\langle \bfq - \bfk, \bfk | H - E_0 | \bfq \rangle - \epsilon_0(\bfq)\   \leftidx{^0}\langle \bfq - \bfk , \bfk | \bfq \rangle,
	\label{eq:3phonon}
\eea
In the second term, we have used the leading order energy of the single-excitation state; this three-excitation vertex will itself correct the single-excitation energy at higher order in perturbation theory. 
The first term in \Eq{eq:3phonon} can be computed directly with the basis states in the previous section:
\bea
	\leftidx{^0}\langle \bfq - \bfk, \bfk | \delta H | \bfq \rangle = \frac{1}{\sqrt{n_0^3 S(\bfq-\bfk) S(\bfk) S(\bfq)} }\int d^3\bfr_1 ... d^3\bfr_N \n^*_{\bfq-\bfk} \n^*_{\bfk} \psi_0 (H - E_0) \n_{\bfq} \psi_0.
\eea
Again, $ (H - E_0)$ acts on $\n_{\bfq} \psi_0$, and after integration by parts plus the fact that $\psi_0$ satisfies $(H-E_0)\psi_0=0$, we can show that
\bea
	&\leftidx{^0}\langle \bfq - \bfk, \bfk | \delta H | \bfq \rangle = \bigsum_j \frac{1}{\sqrt{n_0^3 S(\bfq-\bfk) S(\bfk) S(\bfq)} }\int d^3\bfr_1 ... d^3\bfr_N \frac{ (\psi_0)^2}{2\mHe}  \nabla_j \left( \n^*_{\bfq-\bfk}  \n^*_{\bfk} \right) \nabla_j \left(  \n_{\bfq} \right)\\
	&= \bigsum_j \frac{1}{N \sqrt{ n_0 S(\bfq-\bfk) S(\bfk) S(\bfq)} }\int d^3\bfr_1 ... d^3\bfr_N \frac{ (\psi_0)^2}{2\mHe}  \left( -i (\bfq - \bfk) e^{-i (\bfq - \bfk) \cdot \bfr_j}  \n^*_{\bfk} - i (\bfk) e^{-i \bfk \cdot \bfr_j}  \n^*_{\bfq - \bfk} \right) \left( i \bfq e^{i \bfq \cdot \bfr_j} \right). \nonumber
\eea
We rewrite the terms above in terms of the static structure function,
\bea
	 \frac{1}{ \sqrt{N} }  \sum_i \langle \Psi_0 |  e^{-i \bfq \bfr_i}  \n_\bfq  | \Psi_0 \rangle =  \frac{1}{\sqrt{n_0}} \langle \Psi_0 | \n^*_{\bfq} \n_\bfq  | \Psi_0 \rangle = \sqrt{n_0} S(\bfq).
\eea
Using this result, we obtain 
\bea
	\leftidx{^0}\langle \bfq - \bfk, \bfk | H - E_0 | \bfq \rangle  & = &  \frac{  \bfq \cdot (\bfq - \bfk) S(\bfk) + \bfq \cdot \bfk\, S( \bfq - \bfk)}{2\mHe \sqrt{N}\sqrt{ S(\bfq-\bfk) S(\bfk) S(\bfq)} } 	\label{eq:matrixelstep1}
\eea

Next, to directly compute the overlap matrix element $^0\langle \bfq-\bfk, \bfk | \bfq \rangle$ requires some working assumption for the form of the ground state wavefunction. Alternatively, one may estimate for this overlap term with a more indirect method.  
The simplest ansatz which yields the correct long-wavelength behavior and satisfies a certain set of consistency conditions is known as the ``convolution approximation.'' With this ansatz, one finds \cite{JacksonFeenberg,Feenberg}
\begin{align}\label{eq:overlapterm}
	^0\langle \bfq-\bfk, \bfk | \bfq \rangle = \frac{\sqrt{ S(\bfq-\bfk) S(\bfk) S(\bfq)}}{\sqrt{N}},
\end{align}
which we derive in detail in \App{app:overlap}.
It has been shown that using this form gives good agreement with experimental data on neutron scattering. Various improvements to the convolution approximation have been considered (see {\it e.g.}~\cite{PhysRevB.13.3779}), though for our approximate, analytic treatment we choose to keep the simplest possibility. This has the main advantage that the formulae of the final answer are very manageable. In particular, utilizing Eq.~(\ref{eq:3phonon}), the full matrix element is then given by 
\begin{align}\label{eq:finalmatrixelement}
	\langle \bfq - \bfk , \bfk | \delta H | \bfq \rangle &=  \frac{ \bfq \cdot (\bfq - \bfk) S(\bfk) + \bfq \cdot \bfk\, S( \bfq - \bfk) - q^2 S(\bfk) S(\bfq - \bfk)}{2\mHe \sqrt{N}\sqrt{ S(\bfq-\bfk) S(\bfk) S(\bfq)} } 
\end{align}

\begin{figure}[t]
\includegraphics[width=0.85\textwidth]{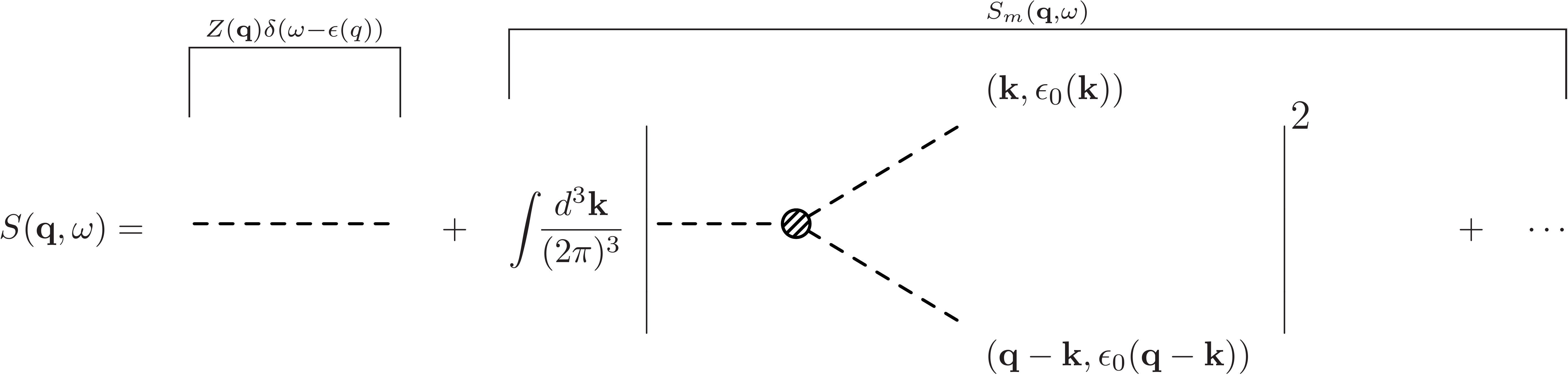}
\caption{Expansion of the dynamic response function in terms of diagrams, where the dashed lines indicate the excitations as defined in \Eq{eq:gramschmidt2}. The two-excitation diagram is the leading contribution to $S(\bfq,\omega)$ for $\bfq,\omega$ away from the dispersion relation. 
\label{Sko_diagrams}}
\end{figure}

Having obtained the three-excitation matrix element, it is now possible to systematically compute the single excitation energy as a perturbation series in this matrix element. To leading order in Brillouin-Wigner perturbation theory \cite{brillouinwignerbook}, the eigenstates of the Hamiltonian are
\bea\label{eq:leadingeigenstates}
|\Psi_\bfq\rangle &=& | \bfq \rangle +\frac{1}{2}\sum_{\bfp,\bfk}|\bfp,\bfk\rangle \frac{\langle \bfp,\bfk|\delta H | \bfq\rangle}{\epsilon(\bfq)-\epsilon_0(\bfk)-\epsilon_0(\bfp)}\delta_{\bfp+\bfk,\bfq}\\
|\Psi_{\bfk,\bfq}\rangle &=& | \bfk,\bfq\rangle +\frac{1}{2}\sum_{\bfp}|\bfp\rangle \frac{\langle \bfp|\delta H | \bfk,\bfq\rangle}{\epsilon(\bfk)+\epsilon(\bfq)-\epsilon_0(\bfp)}\delta_{\bfp,\bfk+\bfq}\,.\label{eq:leadingeigenstates2}
\eea
Similarly, the energy of $|\Psi_\bfq\rangle$ is then given by the recursive relation
\beq
\epsilon(\bfq) =\langle\Psi_\bfq|H|\Psi_\bfq\rangle =\epsilon_0(\bfq) + \frac{1}{2} \int\!\! \frac{d^3 {\bf k}}{ (2\pi)^3}  \frac{V | \langle  \bfq-\bfk,\bfk | \delta H | \bfq \rangle |^2}{ \epsilon(\bfq) - \epsilon_0(\bfq-\bfk) - \epsilon_0(\bfk) }\,.
\eeq
where we took the continuum limit.
To solve for the resummed energy to lowest order, $\epsilon(\bfq)$ is replaced by  $\epsilon_0(\bfq)$ inside the integral above. 
By inserting \Eq{eq:leadingeigenstates2} in \Eq{eq:Sqw}, one can compute the two-excitation contribution to the dynamic structure function to leading order
\begin{align}
 S_m(\bfq, \omega) =  \frac{S(\bfq)}{2} \int \!\!\frac{d^3 {\bf k}}{ (2\pi)^3}  \frac{V | \langle\bfq-\bfk, \bfk | \delta H | \bfq \rangle |^2}{ (\epsilon_0(\bfq) - \omega)^2 }  \ \delta(\omega - \epsilon_0(\bfk) - \epsilon_0(\bfq-\bfk) ).
	\label{eq:Sqw_leadingorder}
\end{align}
This is shown diagrammatically in \Fig{Sko_diagrams}. Whenever we use this approximation, we use the Bijl-Feynman dispersion relation and the measured form of $S(\bfq)$, both shown in \Fig{fig:dispersion}. While this form is enough to obtain a rough estimate of the scattering rate, it clearly has the incorrect structure as it only uses the lowest-order energies. 

\begin{figure}[t]\centering
\includegraphics[width=0.7\textwidth]{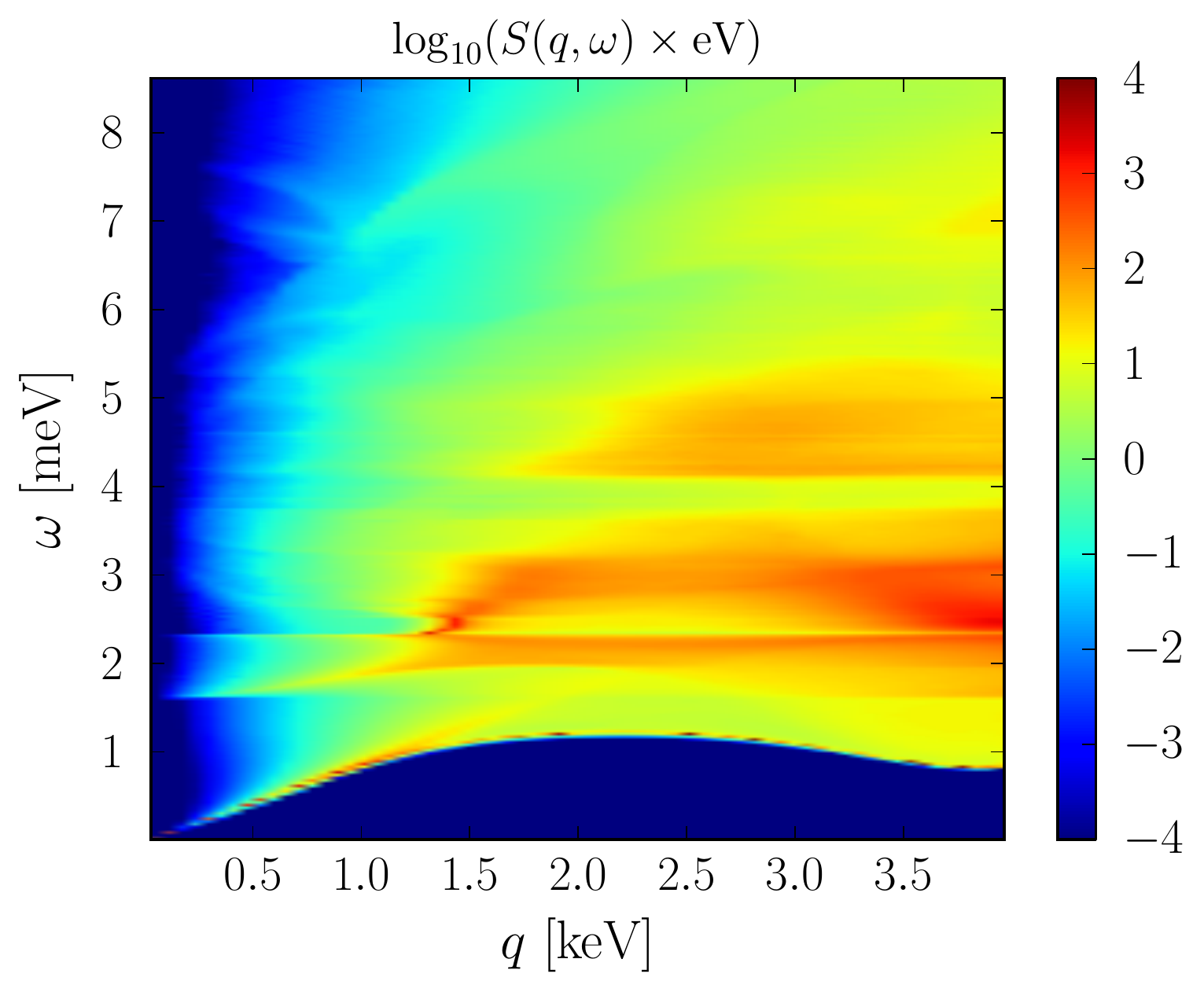}
\caption{  \label{fig:SQw_simulation}  Self-consistent calculation of the dynamic structure function $S(\bfq,\omega)$, obtained from Ref.~\cite{Krotscheck2015} (CKL15). For a given $\bfq$, the onset of the response at a minimum $\omega$ clearly shows the one-excitation component of $S(\bfq,\omega)$. The response at larger $\omega$ corresponds to the multi-excitation component, where the structures at 2 meV and above arise from multi-excitations of rotons/maxons. In the experimental data these  structures are less prominent, which is expected once additional interactions are included (see figures 21-22 and discussion in Ref.~\cite{Krotscheck2015}.)}
\end{figure}

This deficiency is addressed in many detailed calculations of $S(\bfq,\omega)$ found in the literature \cite{manousakis1986theoretical,andersenStirling94,Krotscheck2015}, using different approximations for the three-excitation vertex and in defining the multi-excitation states.  We note that the approach presented here is not entirely unique in giving reasonable agreement with the data. Recently, a fully self-consistent calculation which resums the corrections due to the three-phonon vertex and gives good agreement with experimental data has been presented by Campbell, Krotscheck and Lichtenegger in Ref.~\cite{Krotscheck2015} (hereafter, CKL15).  Rather than model the effect of the interactions with a heuristic ansatz for the overlap term, they explicitly include the leading term from the potential in \Eq{eq:hamfeenberg}. 
Operationally, they obtain $S(\bfq,\omega)$ by recursively solving for the self-energy $\Sigma(\bfq,\omega)$, which satisfies
\begin{align}
	\Sigma(\bfq,\omega) = \epsilon_0(\bfq) +  \frac{1}{2} \int\!\! \frac{d^3 {\bf k}}{ (2\pi)^3}  \frac{V | \langle  \bfq-\bfk,{\bf k} | \delta H | \bfq \rangle |^2}{ \omega- \Sigma(\bfq-\bfk, \omega - \epsilon_0(\bfk) ) - \Sigma(\bfk,\omega-\epsilon_0(\bfq-\bfk) )}.
\end{align}
Using this self-energy, the renormalized energies $\epsilon(\bfq)$ then match the observed single-excitation energies, and the dynamic structure factor is given by the optical theorem
\begin{align}
	S(\bfq, \omega) = -\frac{1}{\pi} \frac{ S(k) {\rm Im}\, \Sigma(\bfq, \omega)}{ (\omega - \epsilon_0(\bfq))^2 + ({\rm Im}\, \Sigma(\bfq, \omega))^2 }.
\end{align}
The result for $S(\bfq,\omega)$ is shown in \Fig{fig:SQw_simulation}, which includes both the single and multi-excitation response.
We emphasize that the method of Ref.~\cite{Krotscheck2015} includes multi-excitation production beyond just the leading order two-excitation production, with the limitation that the multi-excitation production still relies on the three-excitation vertex (in general, higher-point vertices are present).
A detailed comparison of this theoretical calculation with inelastic neutron scattering data can be found in Ref.~\cite{Beauvois2016}. Accounting for neutrons that scatter multiple times in the liquid, the data is in reasonably good agreement with theory for the multi-excitation component.

As we will discuss in the following section, the results shown in \Fig{fig:SQw_simulation} are in broad agreement with the lowest order calculation of $S_m(\bfq,\omega)$ using \Eq{eq:Sqw_leadingorder}, although there are significant differences in detailed structure.  Where available, we will therefore use the numerical results of CKL15 to compute DM scattering, and use the lowest order results only as a guide to extending CKL15 to low momentum transfer.

%%%%%%%%%%%%%%%%%%%%%%%%%%%%%%%%%%%
\section{Reach for dark matter scattering}\label{sec:DMscattering}
%%%%%%%%%%%%%%%%%%%%%%%%%%%%%%%%%%%

We now turn to DM detection with an idealized liquid helium detector, applying our knowledge of the dynamic structure function derived in the previous section. A possible concept for this detector has been shown in~\cite{mckinseyslac}: the basic idea is that a scattering event creates quasiparticle excitations, which can propagate to the surface of the liquid. At the liquid-gas interface, the quasiparticle has a high probability to eject a free helium atom via quantum evaporation, followed by calorimetric detection of the helium atom. Furthermore, the evaporation process may give a natural amplification technique (with amplification factors of $\sim$10), and in principle could be applied for single quasiparticle energies as low as $\omega = 0.6$ meV. 

In this section we use the various results for the dynamic structure function $S(\bfq,\omega)$ to obtain the rate for DM scattering. We discuss the derivation of the rate given in~\cite{Schutz:2016tid} in greater detail, considering the expanded calculation of $S(\bfq,\omega)$.  As a benchmark, we will consider a background-free kg-year exposure. For multi-excitation final states, we take a minimum energy of $\omega = 1.2$ meV and energies up to 8.6 meV. This upper value on $\omega$ coincides with the upper cutoff of the numerical results we take from CKL15; furthermore, this energy range constitutes the bulk of the response, and the rate falls off rapidly at higher $\omega$.

The results of this section are applicable to models of dark matter interacting coherently with helium atoms via a new mediator, where we consider both the heavy mediator and light mediator limits. In contrast, in the long wavelength limit the helium atom does {\emph{not}} have a net charge for a mediator such as a hidden photon. We will discuss signals related to the hidden photon in \Sec{sec:DarkPhoton}.

%%%%%%%%%%%%%%%%%%%
\subsection{Preliminaries\label{sec:DMsetup}}
%%%%%%%%%%%%%%%%%%%

The total DM scattering rate per unit target mass is given by
\begin{align}
	 \frac{dR}{d\omega} & = \frac{1}{\rho_{\mathrm{He}}} \frac{\rho_X}{m_X} \int d^3 \bfv f(\bfv)   \int_{|p_i - p_f|}^{p_i + p_f} dq\  \frac{d \Gamma}{d\omega dq}, 
\end{align}
where $\frac{d \Gamma}{d\omega dq}$ is the differential scattering rate per incoming DM particle. 
 We denote the initial momentum of the DM $p_i$ and the final momentum $p_f$, with
\bea
	 p_f =m_X \sqrt{\bfv^2 - \frac{2 \omega}{m_X} } \ \ , \ \ p_i = m_X |\bfv|.
\eea
For the velocity distribution for the dark matter, we assume the standard halo model Maxwellian distribution, boosted to the earth's frame:
\begin{align}
	f(\bfv) &= \frac{1}{N(v_0, v_{esc})}  \exp \left[ - \frac{(\bfv + \bfv_e)^2}{v_0^2} \right]   \ \ \Theta( v_{esc} - |\bfv + \bfv_e|), \label{eq:velodist} \\
	&N(v_0, v_{esc}) = \pi^{3/2} v_0^3 \left[ {\rm erf} ( \tfrac{v_{esc}}{v_0} ) - 2\tfrac{v_{esc}}{v_0} \exp\left( -( \tfrac{v_{esc}}{v_0} )^2 \right) \right] 
\end{align}
where $v_0 = 220$ km/s, the escape velocity $v_{esc} = 500$ km/s, and we take the average earth's velocity to be $v_e = 240$ km/s. The normalization factor $N(v_0, v_{esc})$ accounts for the hard cutoff in the distribution at $v_{esc}$. For the local dark matter density, we take $\rho_X= 0.3$ GeV/cm$^3$.
(Note that this velocity distribution differs somewhat from that used in Ref.~\cite{Schutz:2016tid}, with more weight at higher initial velocities. This leads to a factor of few larger scattering rate.)

Analogous to the case for neutron scattering \Eq{eq:potential}, DM in superfluid helium sees the potential
\begin{align}
	V(\bfr) = \frac{2\pi b_X}{m_X} \sum_i \delta^{(3)}( \bfr - \bfr_i) = \frac{2\pi b_X}{m_X}  n(\bfr),
	\label{eq:potentialDM}
\end{align}
where $b_X$ is the DM-helium scattering length. This prescription works both for a light mediator and for a contact operator, where in the former case $b_X$ is momentum dependent. We can then compute the scattering rate with Fermi's golden rule,
\beq
\Gamma=2 \pi \left(\frac{2\pi b_X}{m_X} \right)^2 \int\!\!\frac{d^3 p_f}{(2\pi)^3} \sum_\beta  | \langle \Psi_\beta | n_\bfq | \Psi_0\rangle|^2 \delta (E_i-E_f-\omega_\beta).
\eeq
With a suitable change of variables, the differential rate is then 
\beq
\frac{d \Gamma}{d q d\omega} = \frac{1}{2}  \n_0 \frac{\sigma_X(\bfq)}{m_X}\frac{q}{p_i} S(\bfq,\omega),
\eeq
where we used the definition of $S(\bfq, \omega)$ in \Eq{eq:dynamicstructure}. (The derivation of the neutron scattering rate in \Eq{eqn:neutronscattering} is completely analogous.)
The DM-nucleus scattering cross section $\sigma_X (\bfq)=4\pi b_X (\bfq)^2$, where $b_X(\bfq)$ is the DM scattering length. Assuming the DM-nucleus interaction is mediated by a new force carrier $\phi$, we can express this as 
\bea
	\sigma_X(\bfq) \equiv \begin{cases}
	\sigma_p  \frac{ (f_p Z + f_n (A-Z))^2}{f_p^2} , &   m_\phi \gg q \hspace{1cm}  \textrm{(massive mediator)}\\
	\frac{\sigma_p q_{\textrm{ref}}^4}{q^4}  \frac{ (f_p Z + f_n (A-Z))^2}{f_p^2} , &   m_\phi \ll q \hspace{1cm}  \textrm{(massless mediator)}
	\end{cases},
	\label{eq:xsec}
\eea
where we consider the massive and massless mediator limits, and $\sigma_p$ is the DM-proton cross section at a reference momentum transfer $q_{\textrm{ref}} \equiv m_X v_0$. In what follows we take $f_n = f_p$. 
The expression for DM scattering rate is then
\bea\label{eq:ratefinal}
	 \frac{dR}{d\omega} = \frac{\rho_X}{2\mHe m_X^2} \int d^3 \bfv f(\bfv)  \int_{|p_i - p_f|}^{p_i + p_f} dq \frac{q}{ p_i } \sigma_X(\bfq) S(\bfq, \omega)  .
\eea

%%%%%%%%%%%%%%%%
\subsection{Scattering rate and reach}
%%%%%%%%%%%%%%%%

Since a full, self-consistent calculation of $S(\bfq,\omega)$ has been made available in CKL15, we would like to use these results. However, for scattering of light dark matter, the kinematic regime is somewhat different from that of neutron scattering measurements ($q\gtrsim$ keV) and existing simulation data ($q\gtrsim100$ eV). In particular, for dark matter in the keV to MeV range, we expect typical momentum transfer and energy deposits given by
\beq
\mathrm{eV}\lesssim|\bfq|\lesssim\mathrm{keV}\quad\mathrm{and} \quad\mathrm{meV}\lesssim\omega\lesssim\mathrm{eV}.
\eeq
This is partially outside the regime that was considered in CKL15 and is shown in \Fig{fig:SQw_simulation}, which includes $q\approx$ 100 eV - 4 keV and $\omega < 8.6$ meV.
The reason for the kinematic mismatch between dark matter and the data is the relatively large velocity of the dark matter compared to the speed of sound in helium, which pushes the interaction away from the linear dispersion phonon regime.

%%%%%%%%%%%%%
\begin{figure}[p]\centering
\includegraphics[width=0.48\textwidth]{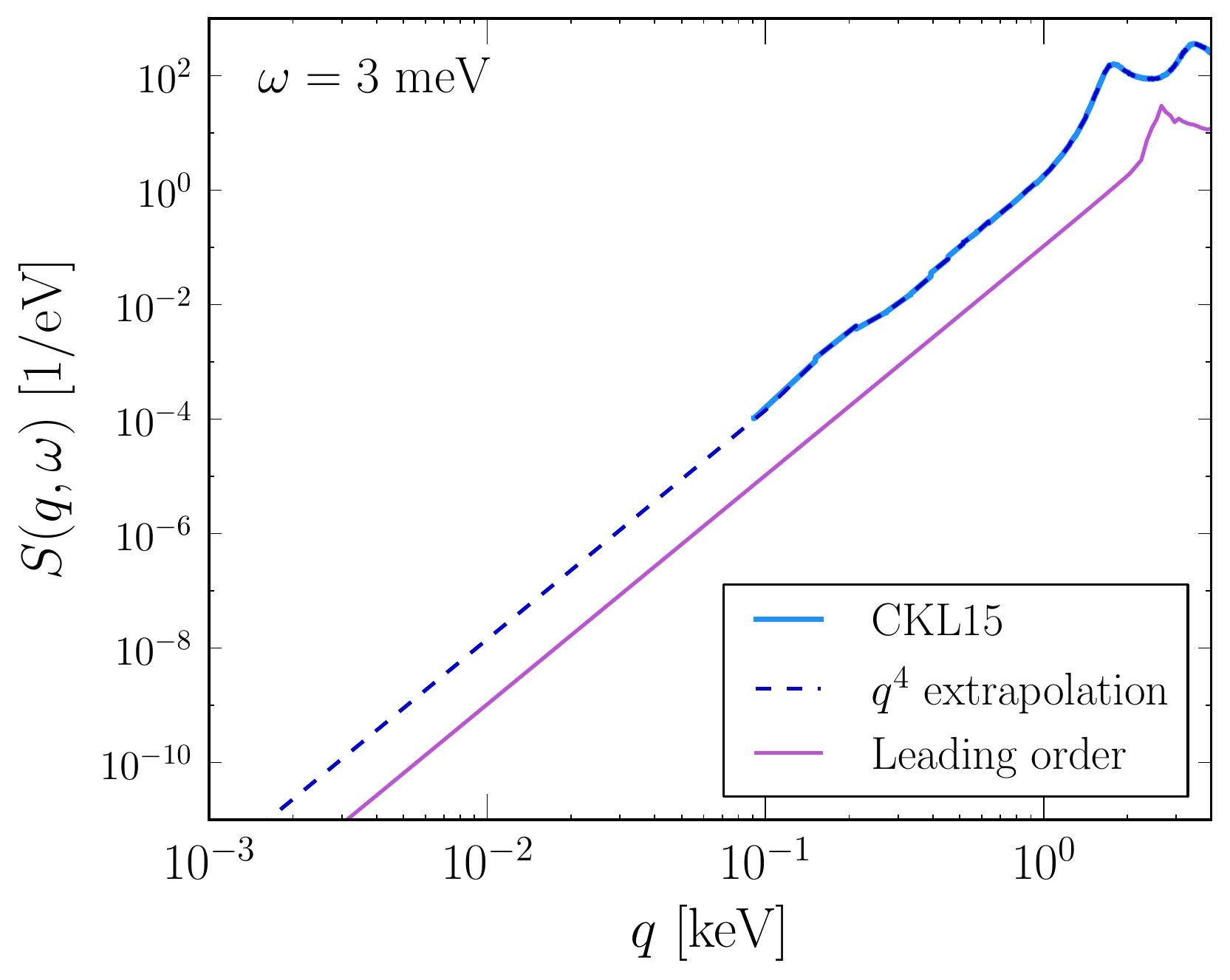}
\hspace{0.12cm}
\includegraphics[width=0.48\textwidth]{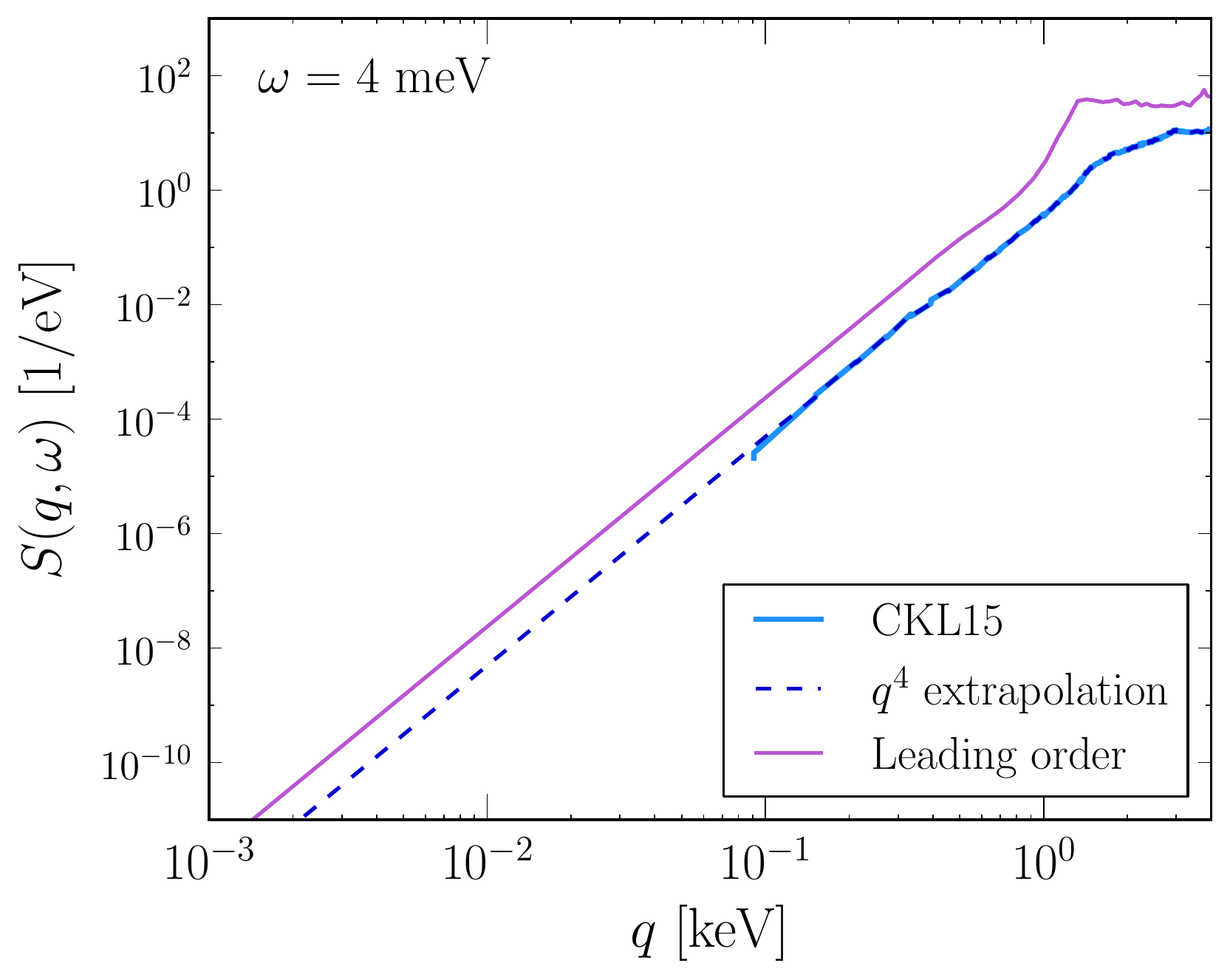}
\caption{
The numerical results from CKL15 are compared with our leading order calculation of $S(\bfq,\omega)$ at two representative values of $\omega$. The dashed line shows the extrapolation with the $q^4$ power law, which is a good fit at low $q$ and agrees with the scaling we find in the leading order calculations.
 \label{fig:k4extrapolation}  }
\end{figure}
%%%%%%%%%%%%%

For the time being, we must therefore rely on a theoretically sensible extrapolation to compute the rate for lighter DM. The numerical data in particular shows a $q^4$ scaling in the low $\bfq$ region, which we can exploit to extrapolate to lower momenta.   This $q^4$ power law can be understood analytically using our approximate expression for $S(\bfq,\omega)$ in \Eq{eq:Sqw_leadingorder}.  In the long-wavelength limit ($q \lesssim$ keV) and at deposited energies  $\omega \gtrsim 0.6 $\, meV, we can take the $q \ll k, |\bfq-\bfk|$ where $\bfk$ and $\bfq-\bfk$ are the momenta of the final state phonons. The matrix element in \Eq{eq:finalmatrixelement} then simplifies to 
\begin{align}
	\langle \bfq - \bfk , \bfk | \delta H | \bfq \rangle &\approx \frac{1}{2\mHe \sqrt{N}} \frac{\bfq^2}{ \sqrt{  S(\bfq)} } \big( 1- S(\bfk ) \big).
\end{align} 
Inserting this in \Eq{eq:Sqw_leadingorder} gives 
\begin{align}
 S(\bfq, \omega) \approx  \frac{1}{16\pi^2}\frac{\bfq^4}{\n_0\mHe^2 \omega^2}\sum_i \tilde\bfk^2_i \big(1-S(\tilde\bfk_i)\big)^2,\
	\label{eq:Sqw_leadingorderapprox}
\end{align}
where the $\tilde\bfk_i$ are the solutions to $\epsilon_0(\bfk_i)=\omega/2$.
We show the $\bfq$-dependence of the numerical data from CKL15 in \Fig{fig:k4extrapolation}, along with the extrapolation to lower $\bfq$ with the $q^4$ power law.  For comparison,  we also show our own numerical calculations of the leading order $S(\bfq,\omega)$ using \Eq{eq:Sqw_leadingorder}, where we took the Bijl-Feynman dispersion relation and measured form of $S(\bfq)$, each shown in \Fig{fig:dispersion}.  (While the Bijl-Feynman dispersion relation is strictly speaking not correct for high momenta, we use it to roughly estimate the contribution from the response above 2 meV, as seen in \Fig{fig:SQw_simulation}.) In both cases, we see the low-$\bfq$ behavior is very well described by a $q^4$ power law.

%%%%%%%%%%%%%
\begin{figure}[p]\centering
\includegraphics[width=0.48\textwidth]{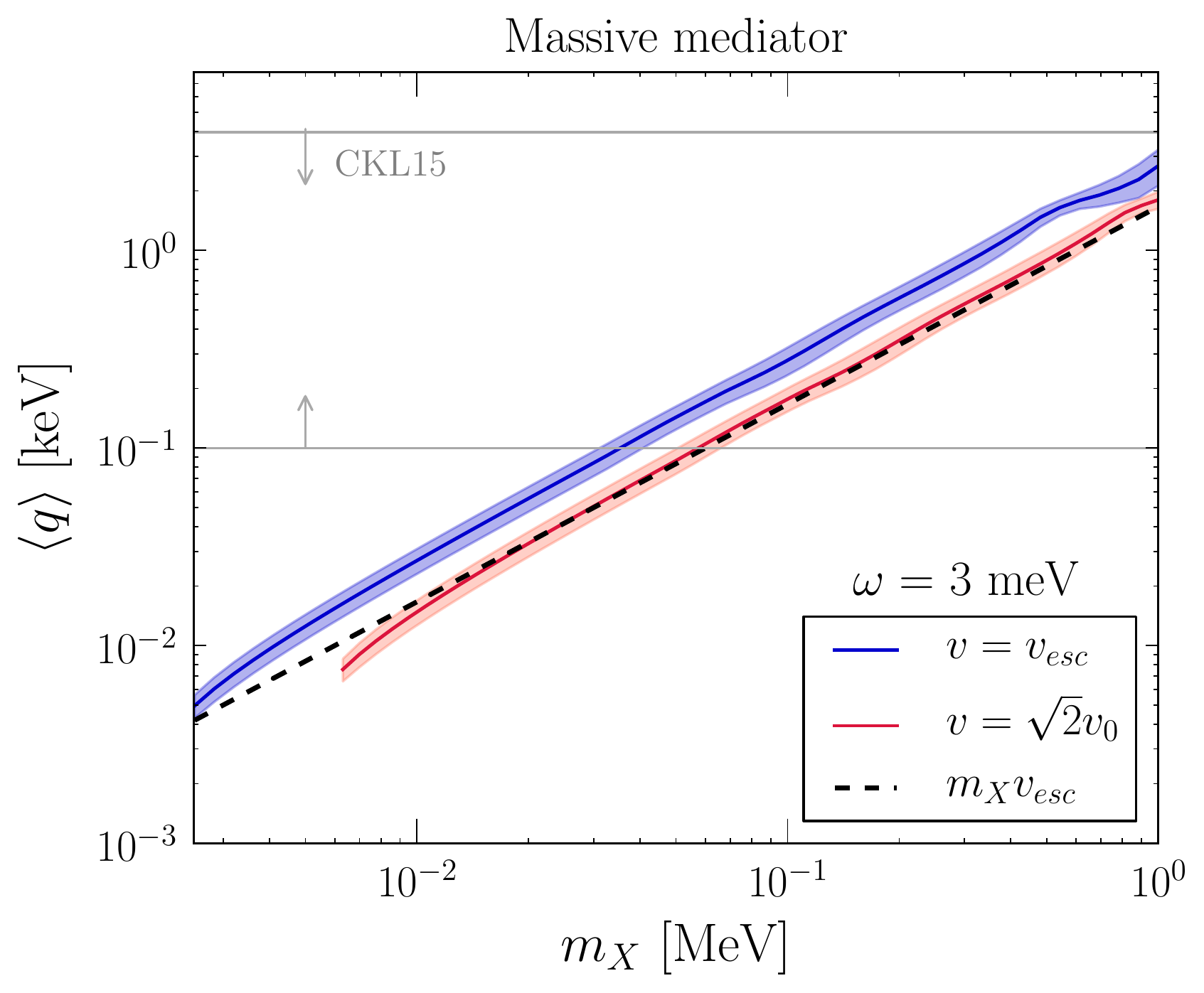}
\hspace{0.12cm}
\includegraphics[width=0.48\textwidth]{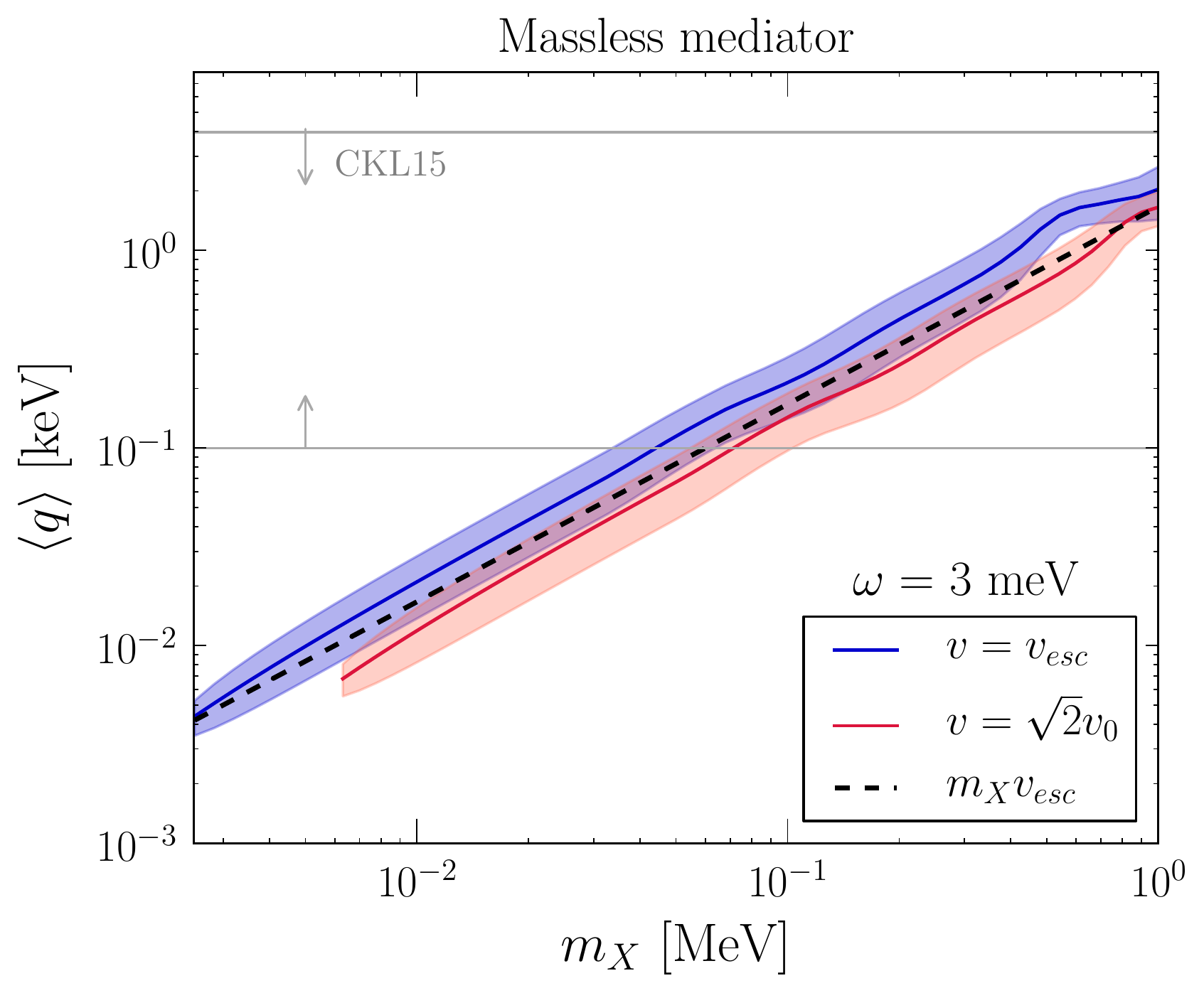}
\caption{
  We show typical values for the total momentum transfer $q = | {\bf q}|$ as a function of dark matter mass $m_X$, considering both a massive mediator ({\bf left}) and massless mediator ({\bf right}). We use $S(\bfq,\omega)$ extrapolated as $q^4$ to plot $\langle q \rangle$  as well as the variance for $q$ (indicated by the shaded region). The energy deposited is fixed at  $\omega = 3$ meV, and we consider two values of the initial DM velocity.  The range of $q$ covered in the CKL15 results (Ref.~\cite{Krotscheck2015}) is indicated by the light gray lines; as can be seen, these numerical results start to be insufficient for DM masses below $\sim$ 50 keV, and we must rely entirely on the $q^4$ extrapolation of the CKL results for masses below $\sim$ 30 keV. 
 \label{fig:Qvals}  }
\end{figure}
%%%%%%%%%%%%%

\begin{figure}[t]\centering
\includegraphics[width=0.48\textwidth]{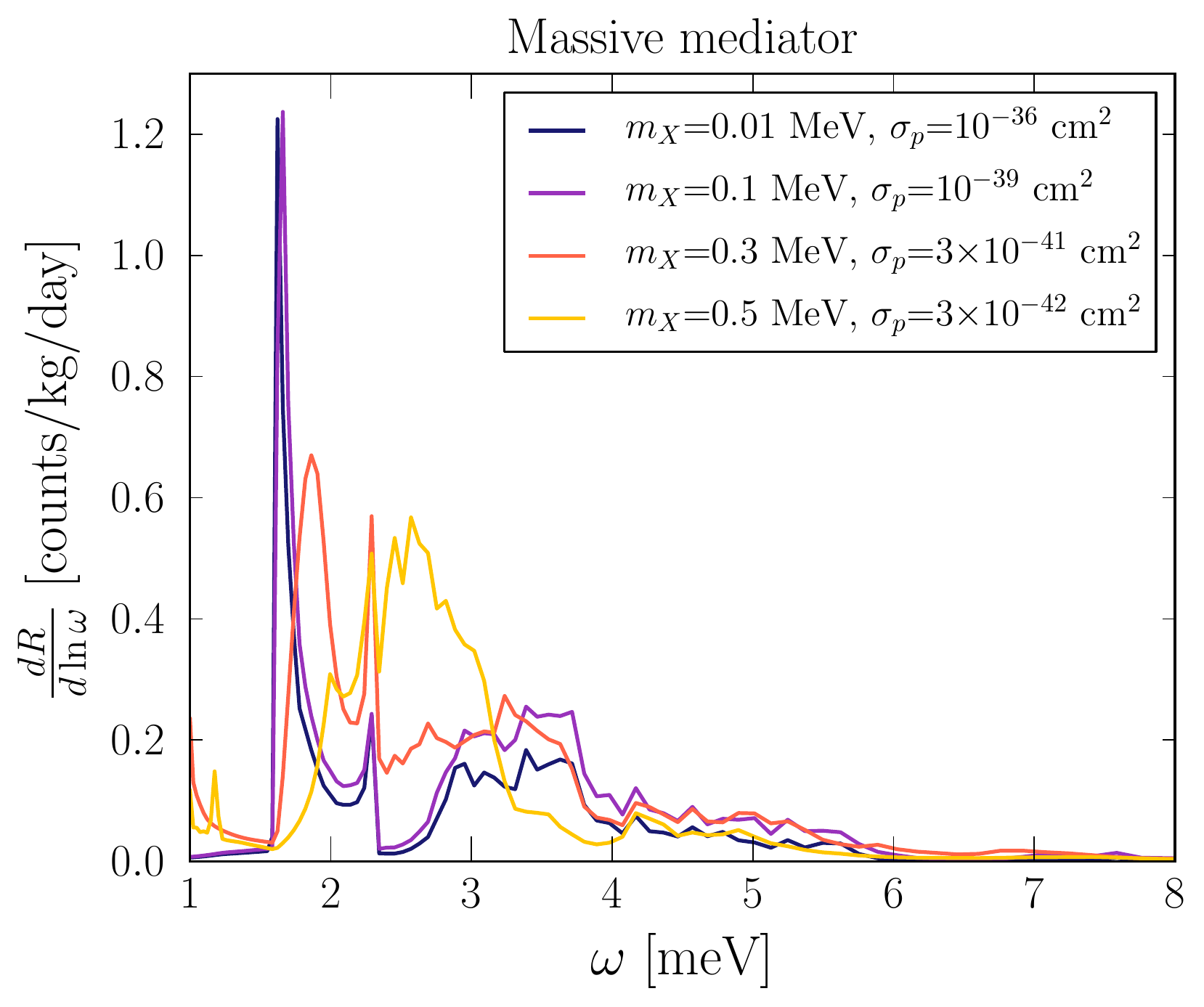}
\hspace{0.12cm}
\includegraphics[width=0.48\textwidth]{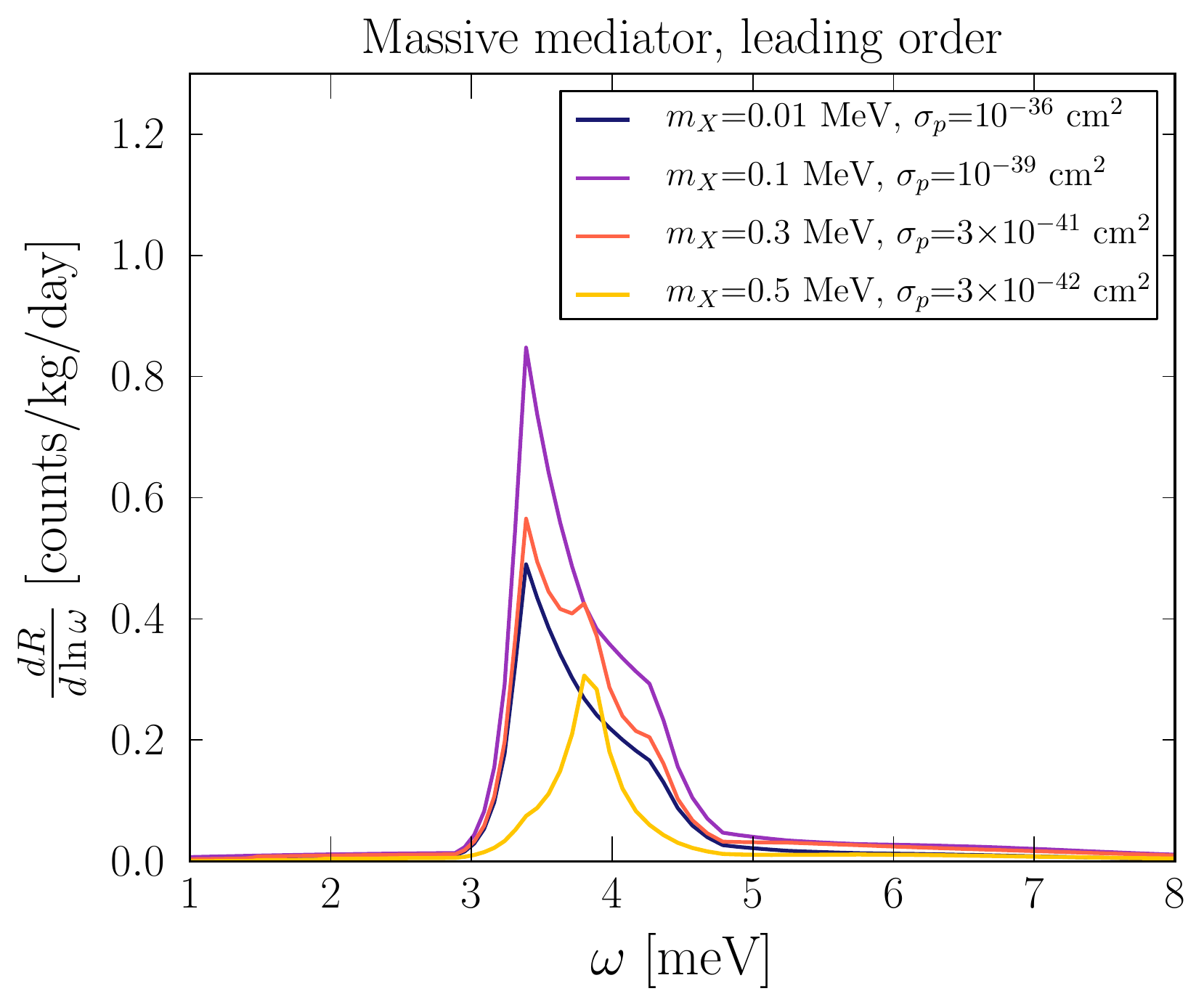}
\caption{  \label{fig:rate_massive}  ({\bf left}) The DM scattering rate via a massive mediator is computed using the $S(\bfq,\omega)$ obtained from CKL15.  ({\bf right}) Here we used the leading order result in \Eq{eq:Sqw_leadingorder}, with the Bijl-Feynman dispersion for single-excitations. There are significant differences in the structure of the spectrum between the two methods, due to the incorrect energies given by the Bijl-Feynman dispersion. However, we find the total integrated rate is similar to within a factor of 2. }
\end{figure}

\begin{figure}[t]\centering
\includegraphics[width=0.48\textwidth]{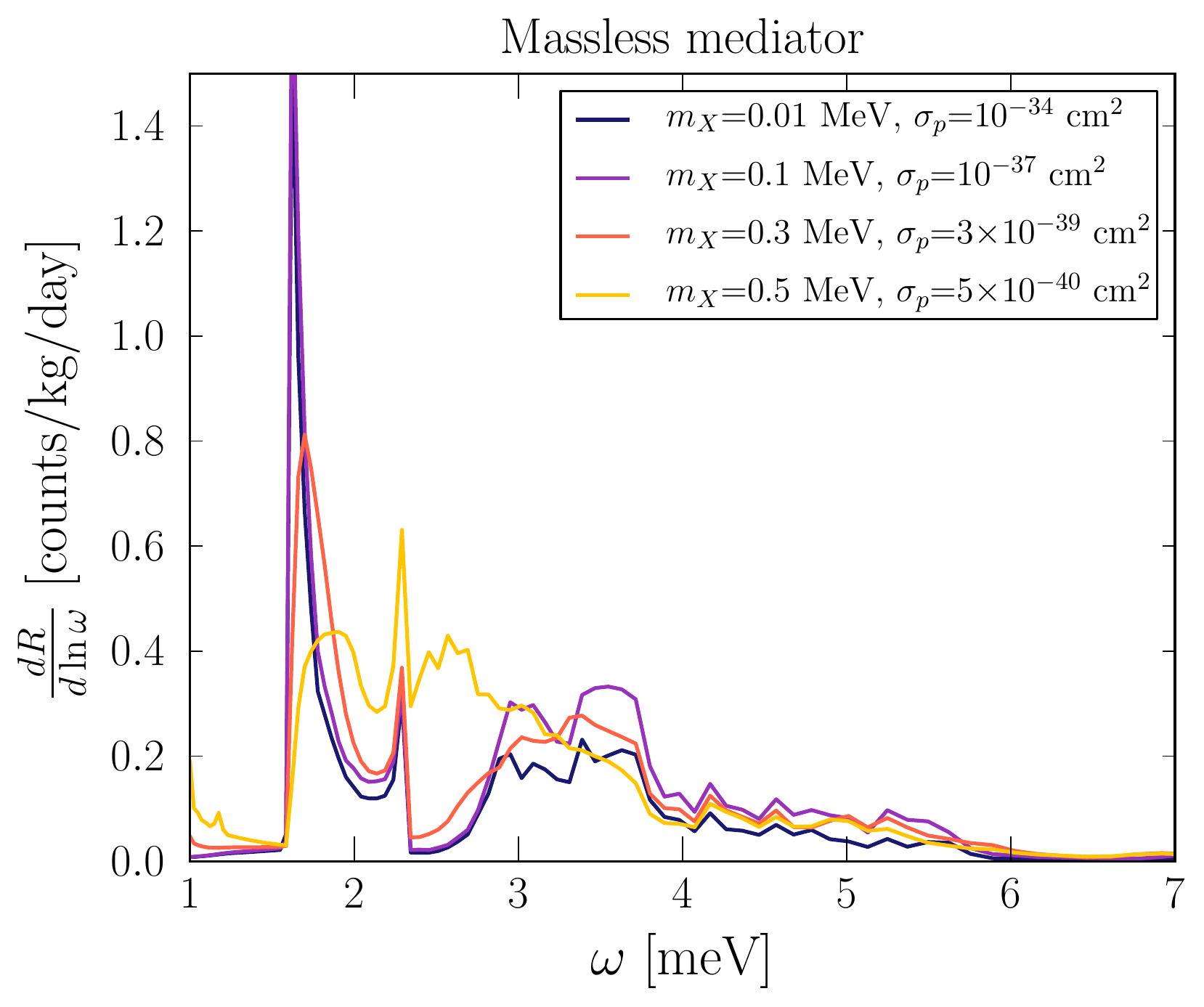}
\hspace{0.12cm}
\includegraphics[width=0.48\textwidth]{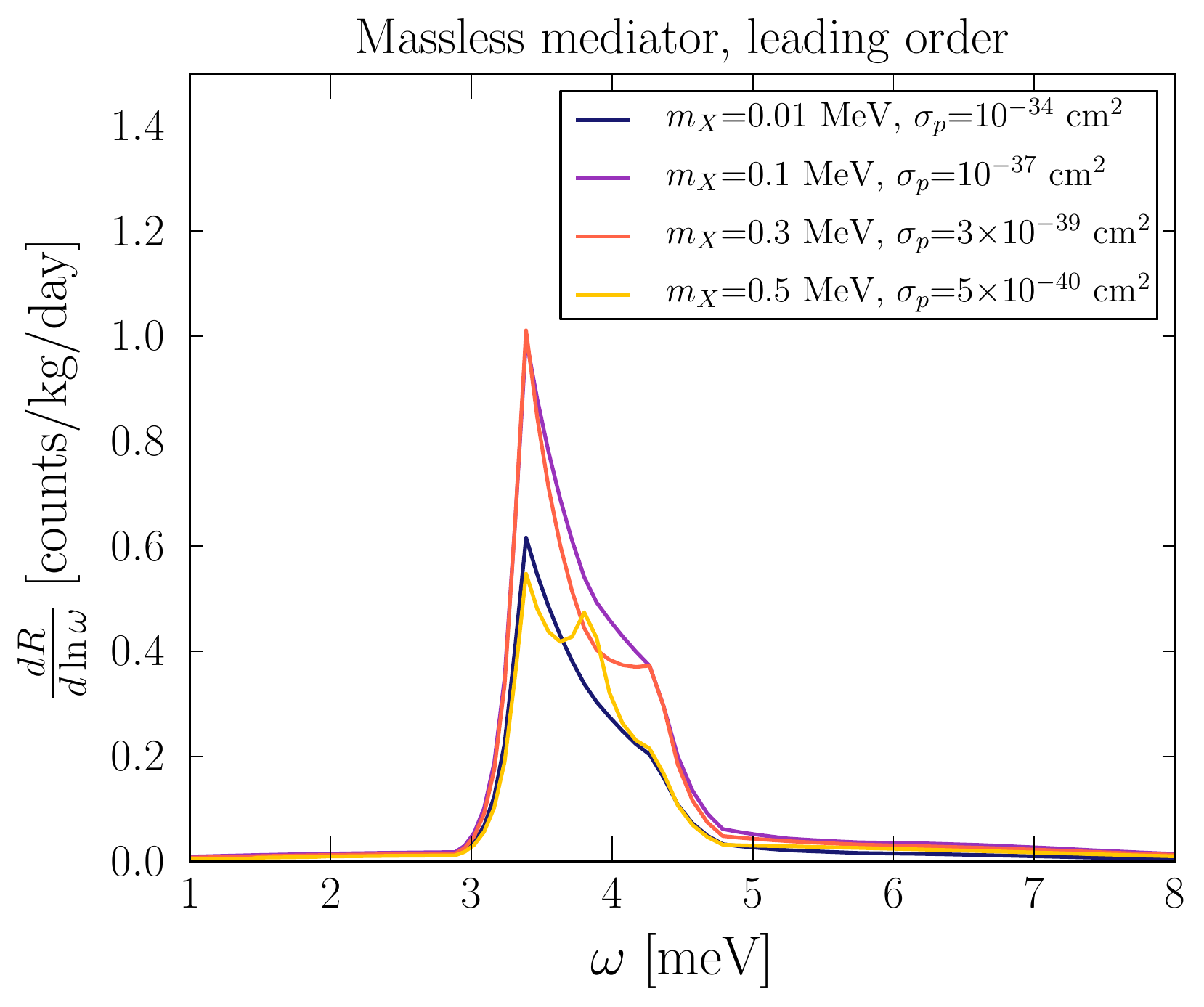}
\caption{  Same as \Fig{fig:rate_massive}, but for DM scattering via a massless mediator. \label{fig:rate_massless} 
}  
\end{figure}

To indicate the relative importance of this extrapolation for dark matter scattering, we show the $\bfq$ values that are most relevant for the DM scattering rate  in \Fig{fig:Qvals}, compared to the momenta covered by the CKL15 results. What is shown in the average $q$, weighted by the relevant factors in \Eq{eq:ratefinal}, or more explicitly
\begin{equation}
\langle q \rangle \equiv \int_{|p_i - p_f|}^{p_i + p_f}\!\!\! dq\, q^2 \sigma_X(\bfq) S(\bfq, \omega) \Bigg/ \int_{|p_i - p_f|}^{p_i + p_f}\!\!\! dq\, q \,\sigma_X(\bfq) S(\bfq, \omega) 
\end{equation}
 using the $S(\bfq, \omega)$ extrapolated below $q=100$ eV with the $q^4$ power law. Thus the DM rate computed here relies heavily on the $q^4$ extrapolation for DM masses below 50-100 keV, and a dedicated simulation along the lines of CKL15 will eventually be needed in this part of parameter space.

\Fig{fig:rate_massive} and \Fig{fig:rate_massless} show the spectrum for scattering via a massive and massless mediator, respectively.  In both cases, we compare the result using $S(\bfq,\omega)$ from CKL15  and that using  Eq.~(\ref{eq:Sqw_leadingorder}). When computing $S(\bfq,\omega)$ from Eq.~(\ref{eq:Sqw_leadingorder}), we use the Bijl-Feynman dispersion for excitations, along with the measured $S(\bfq)$; since this method gives roton/maxon energies which are too high compared to the measured $\epsilon(\bfq)$, the structure here is shifted to higher $\omega$. These differences illustrate the importance of obtaining the correct energies (and widths) of the rotons and maxons, since the rate is clearly dominated by pair-production of these excitations.

\begin{figure}[p]\centering
\includegraphics[width=0.67\textwidth]{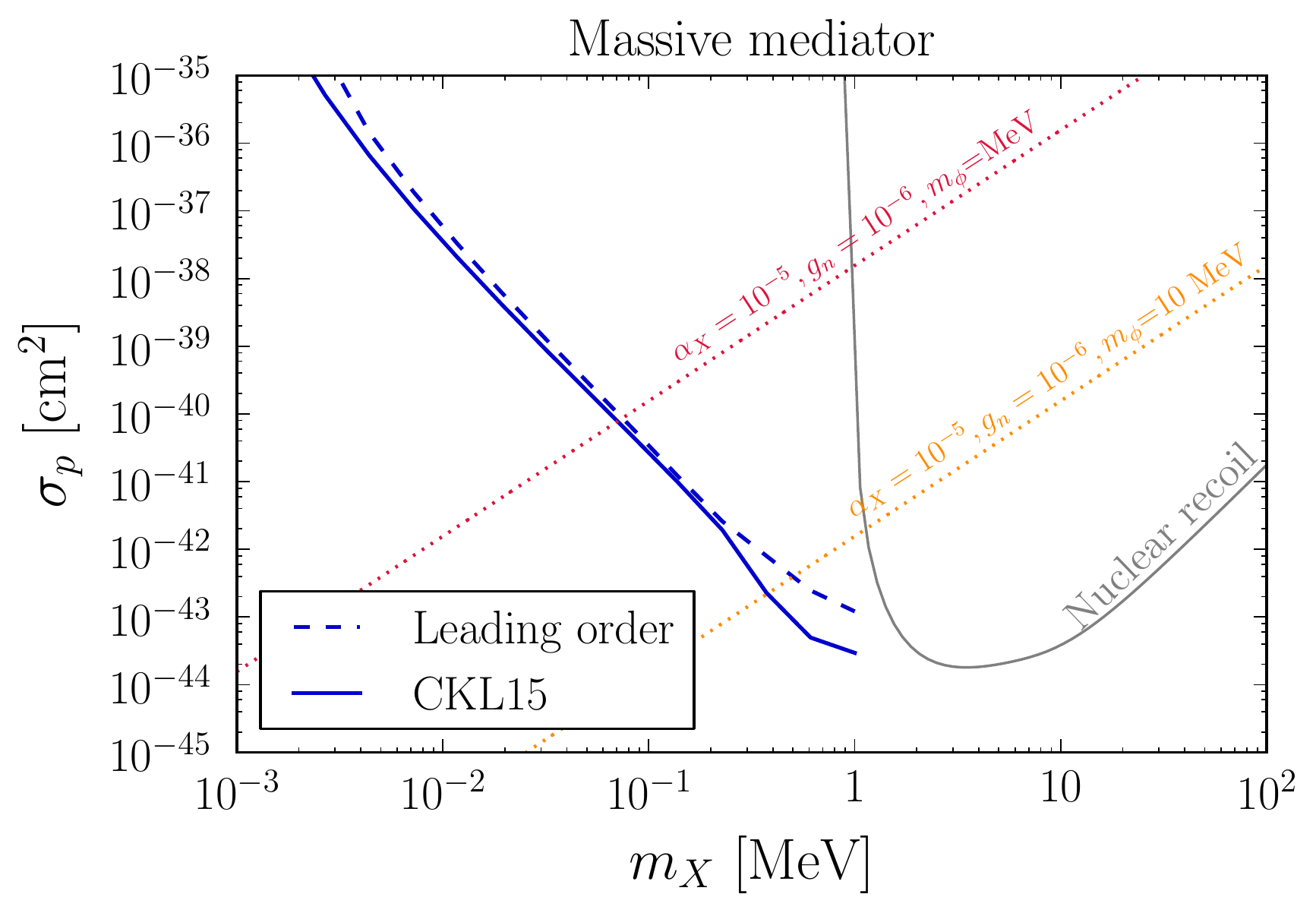}
\vspace{0.2cm}\\
\includegraphics[width=0.67\textwidth]{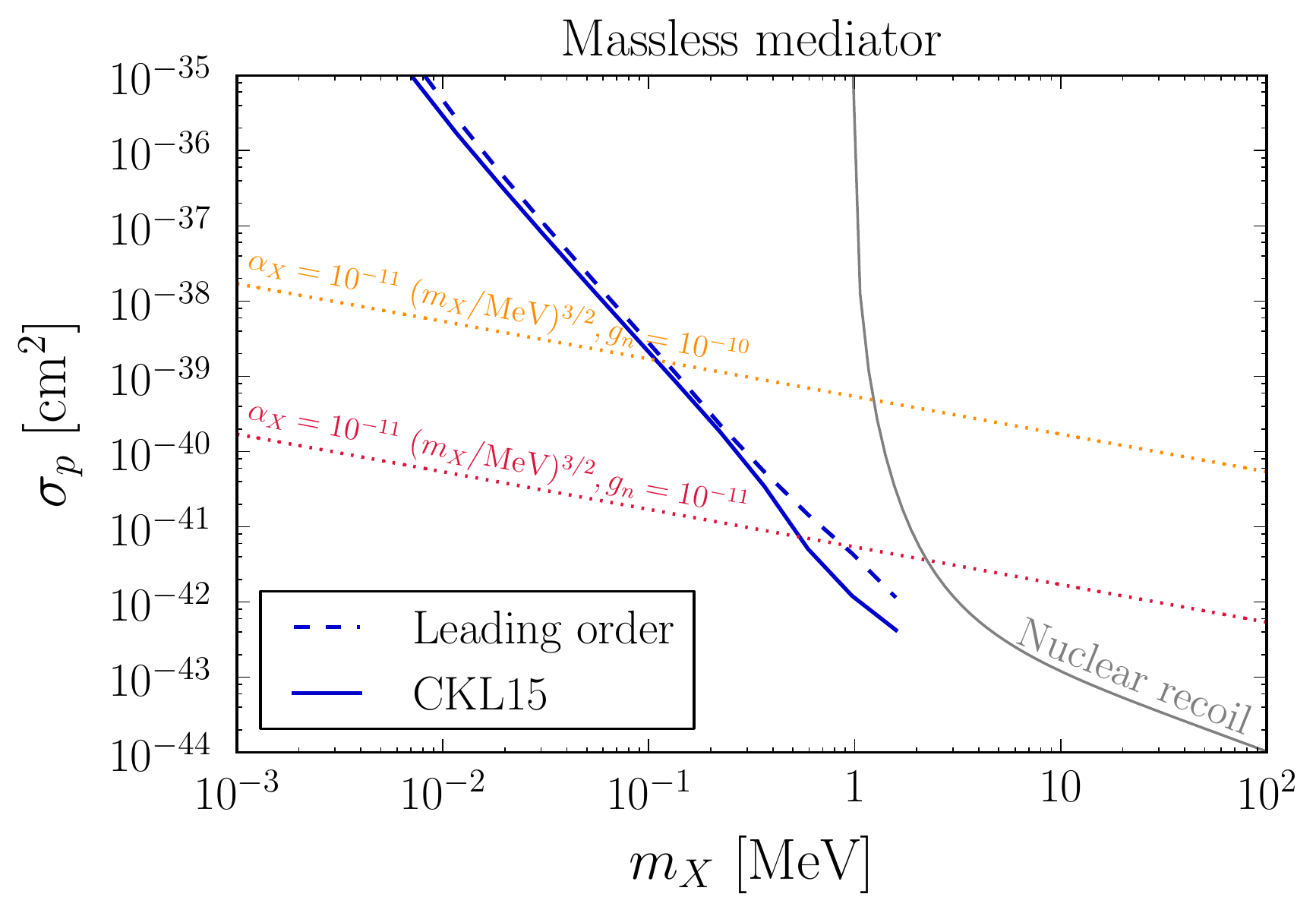}
\caption{Projected reach at 90$\%$ CL (2.4 events) for DM scattering through multi-excitation production in superfluid helium for a 1 kg-year exposure, for the massive mediator ({\bf top}) and massless mediator ({\bf bottom}) cases defined in \Eq{eq:xsec}. The dashed (solid) blue line shows the result using the leading order (CKL15) result for $S(\bfq, \omega)$. We assume zero background and experimental sensitivity down to $\omega \sim $ meV. The reach is derived from the integrated rate with $\omega \in [1.2-8.6]$ meV, where the multi-excitation scattering rate is largest. The reach from ordinary  nuclear recoils is also shown, assuming sensitivity to the energy range $\omega \in [3-100]$ meV (for smaller $\omega$, ordinary nuclear recoils are not possible). The dotted lines show $\sigma_p$ for sample mediator masses and couplings, chosen to roughly satisfy self-interaction, neutron scattering, and stellar bounds (see text).} 
\label{fig:scatter_reach}
\end{figure}

The projected best-case sensitivity for DM scattering is shown in \Fig{fig:scatter_reach}, for 1 kg-year exposure and assuming zero background events.  The results for both computations of  $S(\bfq,\omega)$ are similar once the rate is integrated over the energy range $\omega \in [1.2,8.6]$ meV, despite the significant differences in the spectrum.
In the same plots, we show the reach if only regular nuclear recoils can be observed down to $\sim 3$ meV (gray line).  (Below $\sim 3$ meV, we know that the only modes available are quasiparticle (phonon or roton/maxon) modes -- see \Fig{fig:dispersion}.)  In our estimates, we did not include possible backgrounds from scattering of solar neutrinos (see for example Ref.~\cite{Hochberg:2015fth}) and coherent photon scattering \cite{2016arXiv161007656R}, which are small for these exposures.

 Note our results are consistent with the reach computed in Ref.~\cite{Schutz:2016tid}, where the results utilizing the CKL15 $S(\bfq,\omega)$ match exactly (up to the different velocity distributions used). Ref.~\cite{Schutz:2016tid} also calculated the multi-excitation rate in the leading order approximation, but using a different form of $S(k) = k/\sqrt{ 4 \mHe^2 c_s^2 + k^2}.$ This assumption made it tractable to obtain an analytic result for the rate,    but does not include the peaked spectrum from the rotons that we see in \Fig{fig:rate_massive} and \Fig{fig:rate_massless}. However, accounting for a missing symmetry factor of $1/2$ in the analytic results of Ref.~\cite{Schutz:2016tid} and the different velocity distributions, the reach is similar.

In \Fig{fig:scatter_reach}, we also show contours in $\sigma_p$ for various model-dependent coupling and mass parameters. In particular, the cross section for DM scattering off a single proton or nucleon can be written in the massive and massless mediator limits as
\begin{equation}
	\sigma_p =  \begin{cases}
	 \frac{4 \alpha_X g_n^2 \mu_{n X}^2}{m_\phi^4} , &   m_\phi \gg q_{\rm ref} \hspace{1cm}  \textrm{(massive mediator)}\\
	\frac{4 \alpha_X g_n^2 \mu_{n X}^2}{q_{\rm ref}^4}  , &   m_\phi \ll q_{\rm ref} \hspace{1cm}  \textrm{(massless mediator)}
	\end{cases},
	\label{eq:nucleon_xsec}
\end{equation}
for fixed momentum transfer $q_{\rm ref} = m_X v_0$. Here we have written the mediator coupling to the DM and nucleons as $g_X$ and $g_n$, respectively.  (To relate results with the form of the scattering potential given in \Eq{eq:potentialDM}, we take $b_X/m_X \to A(g_n g_X)/m_\phi^2$ with $A=4$, in the limit of $m_X \ll \mHe$.)

Setting aside the cosmological production mechanism for the DM, there are a number of model-dependent existing constraints on light dark matter, in particular for the case of a light mediator.
The DM-mediator coupling $g_X$ is bounded from DM self-interactions, which can affect DM halo shapes and small-scale structure. The momentum-transfer weighted self-interaction cross section is given by~\cite{Tulin:2012wi},
\begin{align}
	\sigma_T \approx \begin{cases}
	 	\frac{4 \pi \alpha_X^2 m_X^2}{m_\phi^4} , &   m_\phi \gg m_X v \hspace{1cm}  \textrm{(massive mediator)}\\
		\frac{ 16 \pi \alpha_X^2}{m_X^2 v^4} {\rm ln} \frac{m_X v^2}{2 m_\phi \alpha_X} , &   m_\phi \ll m_X v \hspace{1cm}  \textrm{(massless mediator)}
	\end{cases},
\end{align}
where $v$ is the velocity of the DM and in the above we have assumed $2m_\phi \alpha_X/(m_X v^2)\ll 1$, always valid here.
A comparison of observed structure with simulations that incorporate DM self-interactions leads to upper bounds in the ball park of $\sigma_T/m_X \lesssim 0.1-10\, {\rm cm}^2/{\rm g}$, depending on the system. In particular, observations of dwarf galaxies (with $v \sim 10^{-4}$) allow cross sections as high as $\sigma_T/m_X \approx 10\, {\rm cm}^2/{\rm g}$ \cite{Fry:2015rta,Kaplinghat:2015aga} and comparison with shapes of elliptical galaxies (with $v \sim 10^{-3})$ gives an upper bound of about $\sigma_T/m_X \lesssim 0.1\, {\rm cm}^2/{\rm g}$  \cite{Peter:2012jh}.  However, we emphasize these bounds can vary by up to an order of magnitude depending on the detailed modeling of structure formation. Furthermore, existing simulations have focused on hard-sphere scattering, and the bounds may be modified significantly in the massless mediator case, where the scattering is dominantly in the forward direction, leading to less isotropization than the hard-sphere scattering case for a given interaction cross-section.
Nevertheless, we can use these results to obtain an approximate bound on the DM coupling.  Taking $\sigma_T/m_X < 10\, {\rm cm}^2/{\rm g}$ and setting $v \sim 10^{-4}$, we find
\begin{align}
	\alpha_X \lesssim \begin{cases}
	 	2\times10^{-3} \left( \frac{m_\phi}{\rm MeV} \right)^2\sqrt{\frac{\rm MeV}{m_X} } , &   m_\phi \gg m_X v \hspace{1cm}  \textrm{(massive mediator)}\\
		2\times10^{-12} \left( \frac{m_X}{\rm MeV} \right)^{3/2} , &   m_\phi \ll m_X v \hspace{1cm}  \textrm{(massless mediator)}
	\end{cases}, \label{eq:alpha_self}
\end{align}
where in the light mediator limit we took ${\rm ln} \frac{m_X v^2}{2 m_\phi \alpha_X}  \sim 30$.
Note also that we have assumed $m_\phi \gg \alpha_X m_X$, such that quantum mechanical resonance effects can be neglected~\cite{Tulin:2012wi}.

A new mediator which couples to nucleons is also strongly constrained, for instance by measurements of neutron-nucleus scattering or from stellar cooling constraints. For massive mediators at the MeV scale or heavier, neutron-lead scattering experiments set a constraint of~\cite{Barbieri:1975xy}
\begin{align}
	g_n \lesssim 2 \times 10^{-5} \left( \frac{m_\phi}{\rm MeV} \right)^{2}.
\end{align}
Combining this with the self-interaction constraint in \Eq{eq:alpha_self} gives for $m_\phi = {\rm MeV}$ an upper limit of $\sigma_p \lesssim 10^{-33} {\rm cm}^2 \left( m_X/{\rm MeV} \right)^{3/2}$, which is well above the cross sections considered here. For reference, we show the cross section for several parameter choices satisfying the self-interaction and neutron-lead scattering constraints in the top panel of \Fig{fig:scatter_reach}.

For the light mediator case, there are also strong constraints from energy loss in helium burning stars, which require $g_n \lesssim 4\times 10^{-11}$ for mediator mass $m_\phi \lesssim 10\, \textrm{keV}$~\cite{Raffelt:1996wa}. Combined with the self-interaction constraint given in \Eq{eq:alpha_self}, this would put a strong upper bound on the allowed $\sigma_p$. However, in both cases the limits are model-dependent and may be uncertain.  With this caveat, in \Fig{fig:scatter_reach}, we show $\sigma_p$ for couplings that are roughly consistent with  stellar cooling and self-interactions, where for $\alpha_X$ we include the strong $m_X$-dependence of the bound in \Eq{eq:alpha_self}.

Additionally, we expect that for the cross sections shown here, the effect of DM stopping in the earth can be neglected (see {\it e.g.}~Ref.~\cite{Kavanagh:2016pyr} for a recent detailed analysis of this effect). Assuming an average density of $5.5\; \mathrm{g}/\mathrm{cm}^3$ with a chemical composition of Fe ($32\%$), Si ($30\%$), O ($15\%$), Mg ($14\%$) and S ($3\%$), we estimate the mean free path for DM scattering in the earth to be roughly 9000 km for $\sigma_p\sim 10^{-35}\;\mathrm{cm}^2$. The mean free path is therefore larger than the radius of the earth for all cross sections we consider. Moreover, given that every scattering event would only result in a relatively small energy loss, dark matter stopping in the earth can be safely neglected for the cross sections of interest.

\FloatBarrier

%%%%%%%%%%%%%%%%%%%%%%%%%%%%
\section{Hidden photon processes}\label{sec:DarkPhoton}
%%%%%%%%%%%%%%%%%%%%%%%%%%%%

\newcommand{\cm}{\mathrm{cm}}

\newcommand{\mph}{m_{A'}}

A hidden photon is a well-motivated ingredient of many dark matter models, either as a component of the dark matter itself ({\it e.g.}~\cite{Pospelov:2008jk,Nelson:2011sf,Arias:2012az}) or as a mediator for DM interactions. (For a recent review, see Ref.~\cite{Alexander:2016aln}.) The hidden photon $A'$ couples to standard model fields through the kinetic mixing operator,
\begin{equation}\label{eq:kinmix}
	\mathcal{L}\supset \frac{\kappa}{2} F^{\mu\nu} F'_{\mu\nu}\ \,
\end{equation}
where $\kappa$  is the kinetic mixing parameter and $F_{\mu\nu}$ ($F'_{\mu\nu}$) is the photon (hidden photon) field strength. For a massive hidden photon, this mixing leads to a coupling of the hidden photon with the regular electromagnetic current,  $\kappa e A^\prime_{\mu}  J^\mu_{EM}$, after performing a field redefinition $A_\mu \to A_\mu + \kappa A^\prime_\mu$.
Here, we consider two scenarios: in the first case, a fermionic DM candidate with keV-MeV mass scatters via the hidden photon mediator. Since our expressions only depend on $\kappa\times g_{A'}$, with $g_{A'}$ the DM coupling to the hidden photon, for simplicity we set $g_{A'} = 1$ and quote our results only in terms of $\kappa$.
In addition, we will calculate absorption of sub-eV mass hidden photons in helium, assuming that  the hidden photons  constitute the dark matter.

Since the electric charge of the helium atom is screened at long wavelengths (or small momentum transfers $q \lesssim 1$ keV), the analysis of the previous section no longer applies. Instead, a hidden photon (or photon) couples to the medium by inducing a dipole moment, where the strength of the dipole is determined by the atomic polarizability $\alpha$. Note also that this implies there is negligible difference between the in-medium kinetic mixing and the vacuum kinetic mixing, in contrast to other low-threshold targets like superconductors where in-medium effects substantially affect the rate \cite{Hochberg:2015fth}. 

Our treatment of this coupling via the polarizability will closely follow Ref.~\cite{Fetter1972}, which considered photon scattering in liquid helium.  
First, we obtain the photon coupling with the medium. To leading order, the target medium is treated as a linear dielectric, with an atomic polarizability  $\alpha\approx2\times 10^{-25}\ \cm^3$ (see {\it e.g.}~\cite{Fetter1972}) for helium. The polarization of the medium is given by 
\begin{equation}
	{\bf P}(\bfr)= \alpha\, n(\bfr)\, \bfE(\bfr),
\end{equation}
where $n$ is the number density at helium atoms and $\bfE$ is the total electric field in the medium. The interaction Hamiltonian of the polarization with a radiation field $\bfE_{\gamma}$ is then
\begin{equation}\label{eq:HamiltPdotE}
	H_I = -\frac{1}{2}\int\! d^3 r \, {\bf P}(\bfr)\cdot \bfE_{\gamma}(\bfr).
\end{equation}
If the polarization is solely induced by the incident radiation field, then $ \bfE(\bfr) \approx \bfE_\gamma(\bfr)$. From the coupling to the number density $n(\bfr)$, this interaction allows for photon scattering by creation of excitations in the liquid. When just a single excitation is emitted, this process is known as \emph{Brioullin scattering}.\footnote{Another possibility is \emph{Raman scattering}, where in addition to the final state photon, two back-to-back, high momentum phonons are being emitted. However, the rate for Raman scattering is proportional to $\alpha^2$ and is generally three to four orders of magnitude weaker than Brillouin scattering. We neglect it here, but refer to Ref.~\cite{stephenreview} for a review of both Brillouin and Raman scattering in superfluid helium.} Since the sound speed is much smaller than the speed of light, here the phonon excitation only carries a small fraction of the energy, such that the frequency shift in the outgoing photon is minimal.

%%%%%%%%%%%%%%%%%%%%%%%%%%%%
\begin{figure}[t]\centering
\includegraphics[width=0.4\textwidth]{figures/brioullin_diagram}\hfill
\includegraphics[width=0.4\textwidth]{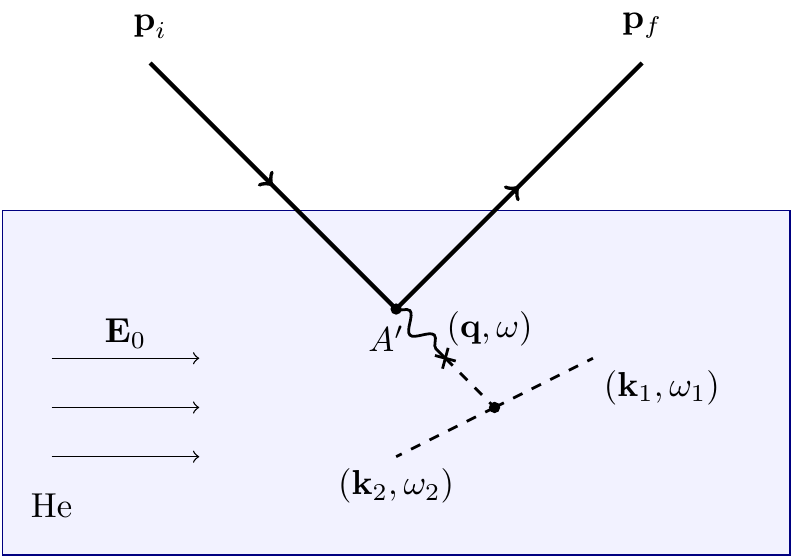}
\caption{  Processes for dark matter scattering  via a hidden photon mediator; the diagrams for absorption of hidden photons are identical to these but without the external fields  $\bfp_i$ and $\bfp_f$.  {\it (left)} In the absence of an external $E$-field, the DM scattering creates a photon and quasiparticle excitation (dashed line) in the final state. The coupling of the hidden photon is given in \Eq{eq:Hdarkphoton}.  {\it (right)} In the presence of an external electric field $E_0$, the intermediate hidden  photon is converted to an off-shell excitation, which subsequently splits into two or more on-shell excitations. See \Eq{eq:HamiltPdotE}. \label{fig:brillouinNoE}
}
\end{figure}
%%%%%%%%%%%%%%%%%%%%%%%%%%%%

To obtain the coupling for the hidden photon field, we perform the field redefinition $A_\mu \to A_\mu + \kappa A^\prime_\mu$, which gives 
\begin{equation}
	\label{eq:Hdarkphoton}
	H_I = - \kappa \alpha\int\!\! d^3 r \,n(\bfr)\, \bfE(\bfr)\cdot \bfE'(\bfr)
\end{equation}
and $\bfE'(\bfr)$ is the hidden photon field. From this, we see that the hidden photon couples to a photon and the density field. A DM scattering (or absorption) would thus give rise to both an observable photon and quasiparticle excitation, as shown in the left panel of \Fig{fig:brillouinNoE}. The physical interpretation is as follows: an incoming hidden photon must first induce a polarization in the medium, which subsequently relaxes back to the ground state by emitting a photon and a phonon. We calculate the rate for these processes in \Sec{sec:withoutE}.

Additionally, the polarization vector ${\bf P}$ may be present already if the experimental setup includes a strong external electric field applied in the liquid. In particular, in neutron EDM experiments, superfluid helium is used for storage of the cold neutrons, and a strong electric field is applied to study the neutron spin precession. Recently, a stable electric field as high as $100$ kV/cm has been demonstrated~\cite{Ito:2015hwa}. The interaction Hamiltonian in this case is
\begin{equation}\label{eq:HamiltPdotE}
	H_I = -\frac{\kappa \alpha}{2}\int\! d^3 r \, n(\bfr)\, \bfE_0 \cdot \bfE'(\bfr),
\end{equation}
where  the external field $ {\bf E_0}$ allows for conversion of the hidden photon into a density perturbation. In this case, there is no final state photon produced, but the  kinematics of light DM scattering requires us to consider the multi-excitation final state, analogous to the discussion in previous sections. This process is shown in the right panel of \Fig{fig:brillouinNoE}, and we calculate the corresponding rates in \Sec{sec:withE}.

%%%%%%%%%%%%%%%%%%%%%%%%%%%%%%%%%%%%%%%%
\subsection{Scattering and absorption without an external E-field} \label{sec:withoutE}
%%%%%%%%%%%%%%%%%%%%%%%%%%%%%%%%%%%%%%%%

We first consider DM scattering in the absence of any external $E$ fields. The process is shown in \Fig{fig:brillouinNoE}, which also defines our conventions for the kinematic variables. For non-relativistic DM, a typical scattering is characterized by a small deposited  energy, but a relatively sizable momentum transfer ($q\sim 10^3 \times \omega$). Since the speed of sound in the superfluid is much smaller than the speed of light, nearly all the deposited energy will be carried away by the photon, while the phonon will absorb the momentum:
\begin{equation}
\omega_1\approx \omega,\quad \bfk_2\approx \bfq\quad \mathrm{and}\quad\omega_2\approx 0.
\label{eq:noE_kinematics}
\end{equation}

To calculate the matrix element, we quantize the electric field of the photon in an arbitrary volume $V$ by
\begin{equation}\label{eq:photonquantization}
 \bfE(\bfr)=\frac{i}{\sqrt{2V}} \bigsum_{\bfk_1,\lambda}\sqrt{\omega_1}\left[ {\bfe}(\bfk_1,\lambda) a_{\bfk_1,\lambda} e^{i \bfk_1\cdot \bfr}
- {\bfe}^\ast(\bfk_1,\lambda) a^\dagger_{\bfk_1,\lambda} e^{-i \bfk_1\cdot \bfr}\right]
\end{equation}
where ${\bfe}(\bfk_1,\lambda)$ is the polarization vector and $a_{\bfk_1,\lambda}$ ($a^\dagger_{\bfk_1,\lambda}$) the annihilation (creation) operators.  Since the DM is non-relativistic,
it can be viewed as sourcing a Coulomb potential for the hidden photon with
\bea
 \bfE'(\bfr) &=& -\mathbf{ \nabla} \Phi'(\bfr)\\
\Phi'(\bfr)&=&\int\!\! d^3\bfr'\frac{:X (\bfr')^\dagger X(\bfr'):}{4\pi| \bfr- \bfr'|}\, e^{- | \bfr- \bfr'| m_{A'} }
\eea
where the $:$ indicates normal ordering and $X$ is the dark matter operator.
After Fourier transforming, the hidden photon field can then be written as
\begin{equation}
 \bfE'(\bfr)=-i\frac{1}{V} \bigsum_{\bfq,s}\; \bfq\; e^{i \bfq\cdot \bfr}\, \frac{N_{X}(\bfq,s) - N_{\bar X}(\bfq,s)}{\bfq^2+m_{A'}^2},
\end{equation}
where $m_{A'}$ is the hidden photon mass and we set the DM charge with respect to the hidden photon equal to one. $N_{X}(\bfq,s)$ and $N_{\bar X}(\bfq,s)$ are the number operators for dark matter and anti-dark matter respectively, where $s$ denotes the spin. Finally, we take the density field $n(\bfr)$ from \Eq{eq:densitydef}.  
In all of the above expressions the momenta $\bfq$, $\bfk_1$ and $\bfk_2$ are summed over, and their naming conventions are arbitrary. However, to make the notation as transparent as possible, we chose to label them according to the external state in \Fig{fig:brillouinNoE} they will eventually contract with.

Using \Eq{eq:Hdarkphoton}, one can obtain the relevant term in the interaction Hamiltonian
\bea
H_I =  - \frac{\alpha \kappa}{V} \bigsum_{ \substack{ \bfk_1,\bfk_2, \bfq \\ s,\lambda} } \sqrt{\frac{\omega_1}{2}}\, \frac{ \bfq\cdot {\bfe}^\ast (\bfk_1,\lambda)}{\bfq^2+{m_{A'}}^2} a_{\bfk_1,\lambda}^\dagger \n_{-\bfk_2} N_{X}(\bfq,s) \, \delta_{\bfq,\bfk_1+\bfk_2}.
\eea
The polarization-averaged squared matrix element is then given by
\bea
|\langle \bfp_i| H_I | \bfp_f; \bfk_1; \bfk_2 \rangle|^2 &=&\frac{\alpha^2\kappa^2}{2V^{2}} \omega_1\frac{\bfq^2}{(\bfq^2+{m_{A'}}^2)^2}  \big|  \langle \Psi_0 |  \n_{-\bfk_2}| \bfk_2 \rangle \big|^2\delta_{\bfq,\bfk_1+\bfk_2}\\
&=&\frac{\alpha^2\kappa^2 \n_0}{2V^{2}} \omega_1\frac{\bfq^2}{(\bfq^2+{m_{A'}}^2)^2}  S(\bfk_2) \delta_{\bfq,\bfk_1+\bfk_2},
\eea
where we used  Eqs.~(\ref{eq:singleexcitation}),~(\ref{eq:Sqdefinition}) in the last step.

The scattering rate is given by Fermi's golden rule,
\bea
\Gamma= 2\pi \sum_{\bfk_1,\bfk_2,\bfp_f} |  \langle \bfp_i| H_I | \bfp_f; \bfk_1; \bfk_2 \rangle |^2 \delta (E_i-\omega_1-\omega_2-E_f)
\eea
with $E_{i,f}$ the initial and final state energy of the dark matter. In the continuum limit, the rate can then be written as 
\begin{align}
\Gamma=&\frac{1}{2} \frac{1}{(2\pi)^5} \alpha^2\kappa^2n_0 \int\!\! d^3\bfp_f\, d^3\bfk_1\,d^3\bfk_2\,  \omega_1\,\frac{\bfq^2}{(\bfq^2+{m_{A'}}^2)^2}  S(\bfk_2)\delta(\omega-\omega_1-\omega_2)\delta^{(3)}(\bfq-\bfk_1-\bfk_2)  \label{eq:GammabrioullinnoE}
\end{align}
If we trade $\bfp_f$ for $\bfq$, eliminate the $\bfk_2$ integral with the momentum $\delta$-function and take  $\omega_2\approx 0$, the integral above can be written in terms of two angles and two magnitudes
\bea
\Gamma &=&n_0 \alpha^2 \kappa^2\frac{1}{2(2\pi)^3}\int\!\! d q\, d \omega\, d\cos\theta\,d\cos\psi   \frac{q^4 \omega^3}{(q^2+m_{A'}^2)^2} \nonumber\\
&&\times S\left(\sqrt{q^2 +\omega^2 - 2 q\,\omega\,\cos\psi}\right)\delta\left(\frac{-q^2 + 2\,  q\,  p_1\cos\theta}{2m_X}-\omega\right)
\eea
where we used the photon dispersion relation to trade $k_1=\omega_1\approx\omega$. Since the final state momentum is $\bfk_2 \approx \bfq \sim m_X v$, for DM masses below an MeV, the momentum transfer is in the linear regime for $S(\bfk_2)$ and we can take $S(\bfk_2) \approx |\bfk_2|/2 \mHe c_s $ in our calculation. We can then evaluate the integrals over the angles and $q$ to obtain the differential rate, which is
\begin{align}
\frac{d \Gamma}{d\omega}= & \frac{n_0 \alpha^2 \kappa^2\omega^3 m_X}{8\pi^3 c_s \mHe p_i} \sqrt{p_i^2-2m_X\omega}
\end{align}
 in the $m_{A'} \ll q$ limit. The total integrated rate is 
\bea
\Gamma (v) \approx\frac{1}{1260\pi^3}\frac{n_0 \alpha^2 \kappa^2 m_X^5 v^8}{ \mHe c_s}.
\eea
The total rate is then 
\begin{align}
	R&=\frac{1}{\rho_{\rm He}}\frac{\rho_{X}}{m_X}\int d^3 \bfv f(\bfv) \Gamma(v)\\
	&\approx 9.2\times 10^{14}\times \kappa^2\times \left(\frac{m_X}{\mathrm{MeV}}\right)^4 / \mathrm{kg}/\mathrm{year}
\end{align}
with $f(\bfv)$ the dark matter velocity distribution in \Eq{eq:velodist}. Given that current stellar constraints on hidden photons already require $\kappa \times m_{A'}\lesssim 3\times 10^{-12}$ eV \cite{An:2013yua,An:2014twa} (for the case of Stueckelberg breaking of the hidden force), the reach in $\kappa$ is not particularly promising.

Next, we consider the scenario where the hidden photon itself is the dark matter, taking $m_{A'}$  to be sub-eV. (For heavier $m_{A'}$ in the eV-keV range, semiconductor targets are a more promising target~\cite{Hochberg:2016sqx,Bloch:2016sjj}.) In this case, the hidden photon can be absorbed by the superfluid, such that $\omega = m_{A'}$, which again results in the emission of a phonon and a real photon with energy $\omega_1 \approx \omega$. 
The computation is analogous to the one outlined above, with the exception that for the hidden photon we must use the expansion analogous to \Eq{eq:photonquantization}. Again using \Eq{eq:Hdarkphoton}, the relevant term in the Hamiltonian is then 
\begin{align}
	H_I &= \alpha \kappa  \frac{1}{2\sqrt{V}} \bigsum_{\bfk_1,\bfk_2,\bfq} \sqrt{\omega_1\omega} \, a_{\bfk_1,\lambda}^\dagger \n_{-\bfk_2} a'_{\bfq,\lambda'} \bfe'(\bfq,\lambda') \cdot  \bfe^\ast(\bfk_1,\lambda) \delta_{\bfq,\bfk_1+\bfk_2}
\end{align}
where $a'_{\lambda,\bfq}$ and $\bfe'(\bfq,\lambda)$ are respectively the hidden photon destruction operator and polarization vector. The momentum transfer in this case is given by $m_{A'} v$, such that we can again take the linear regime for the structure factor, $S(\bfk_2) \approx |\bfk_2|/2 \mHe c_s$. This approximation is always justified for absorption, since $q\ll\omega\lesssim$ eV. The absorption rate can then be obtained with a similar computation to one described above, with
\begin{equation}
	\Gamma =\frac{1}{12\pi} \frac{n_0\alpha^2\kappa^2 m_{A'}^5}{m_{He} c_s}.
\end{equation}
which is independent of the hidden photon velocity. The total observable rate is
\begin{align}
	R&=\frac{1}{\rho_{\rm He}}\frac{\rho_{X}}{m_{A'}}\Gamma\\
	&\approx 7.8 \times 10^{16}\times \kappa^2\times \left(\frac{m_{A'}}{\mathrm{eV}}\right)^4 / \mathrm{kg}/\mathrm{year}.
\end{align}
This rate is only competitive with current stellar constraints on the mixing parameter $\kappa$ if $m_{A'} \sim 1 \mbox{ keV}$. However with an energy deposition as large as 1 keV, other experiments, such as semiconductor targets \cite{Hochberg:2016sqx,Bloch:2016sjj}, are likely to be more sensitive.

%%%%%%%%%%%%%%%%%%%%%%%%%%%%%%%%%%%%%%%%%%%
\subsection{Scattering and absorption with an external E-field} \label{sec:withE}
%%%%%%%%%%%%%%%%%%%%%%%%%%%%%%%%%%%%%%%%%%%

If an external background electric field $\bfE_0$ is turned on, then the medium already has a polarization ${\bf P}_0 =  \alpha\  n(\bfr) \bfE_0$ and the interaction Hamiltonian is given by \Eq{eq:HamiltPdotE}.
 The presence of the external field allows a hidden photon to be converted to a density perturbation, as shown in the right panel of \Fig{fig:brillouinNoE}. As for the case of hard sphere scattering considered in \Sec{sec:DMscattering}, energy and momentum conservation does not allow for a single phonon excitation and the leading process necessarily involves multiple excitations.

For DM scattering, we follow the same treatment of the hidden photon as in the previous section, and we obtain the quantized interaction Hamiltonian,
\bea
H_I =  \frac{\alpha \kappa}{2} \,  \frac{ i}{\sqrt{V}} \bigsum_{\bfq,\bfq',s}  \frac{ \bfE_0\cdot \bfq}{q^2+{m_{A'}}^2} \n_{-\bfq'}\, N_{X}(\bfq,s)\, \delta_{\bfq,\bfq'}
\eea
which we can directly match onto \Eq{eq:potential} by defining an effective dark matter scattering length
\begin{equation}
\frac{2\pi b_{X}}{m_X}= \frac{1}{2}\kappa \frac{ \alpha \bfE_0\cdot \bfq}{q^2+{m_{A'}}^2}.
\end{equation}
With $\sigma_X=4\pi b_X^2$, we can now directly use all the results from \Sec{sec:DMscattering}. 
Interestingly, the rate depends on the direction of the background electric field $\bfE_0$, which in principle induces a daily modulation in the scattering rate. To obtain an upper bound on the rate, we take the field to be parallel with the momentum transfer, after which we obtain
\beq
\frac{d\Gamma}{d q\, d\omega}= \frac{1}{8\pi}\n_0\,m_X  \kappa^2 \alpha^2 \frac{q}{p_i} \frac{  |\bfE_0|^2 q^2}{\left(q^2+{m_{A'}}^2\right)^2} S(\bfq,\omega).
\eeq
For an electric field of $E_0 = 100$ kV/cm, we find that the upper bound on the potential reach for $m_{A'}\ll q$ is
\beq
\kappa\sim 2\times 10^{-9}\times \left(\frac{\mathrm{MeV}}{m_X}\right)^{3/2} 
\eeq
for a kg-year of liquid helium.
This value of $\kappa$ is only competitive with stellar constraints for $m_{A'} \lesssim 10^{-3}$ eV. 

For the case of hidden photon absorption, the Hamiltonian is
\begin{align}
	H_I &=    \frac{\alpha \kappa}{  2 \sqrt{2} } \sum_\bfq \sqrt{\omega}  \,a'_{\bfq,\lambda}\, \n_{-\bfq}\, {\bfe}'(\bfq,\lambda) \cdot \bfE_0.
\end{align}
Next we can compute the polarization averaged, squared matrix element and sum over all multi-excitation final states of the superfluid
\bea
\sum_\beta \big|\langle\bfq  |H_I |\Psi_\beta \rangle\big|^2 &=& \frac{1}{24 }|\bfE_0|^2 \ \alpha^2 \, \kappa^2\omega\sum_\beta \big|\langle \Psi_0 | \n_{-\bfq} | \Psi_\beta\rangle\big|^2.
\eea
Considering multi-excitation production,
\bea
\Gamma&=& \frac{\pi}{12}   | \bfE_0|^2\, \alpha^2\, \kappa^2\, \omega\sum_\beta \big|\langle \Psi_0 | \n_{-\bfq} | \Psi_\beta\rangle\big|^2\delta(\omega_\beta-\omega)\\
&=&  \frac{\pi}{12}  \n_0\ | \bfE_0|^2\, \alpha^2\,\kappa^2\,\omega\, S(\bfq,\omega).
\eea
For hidden photon absorption, the kinematics dictates $q \sim 10^{-3}\times \omega$ with $\omega \lesssim $ eV. For such small momentum transfers and comparatively large energies, we expect a strong suppression of the dynamic structure factor $S(\bfq,\omega)$ due in part to the $q^4$ dependence discussed in \Sec{sec:DMscattering}, as this regime is very far away from the dispersion relations of the quasi-particle states we seek to scatter off. In particular, from \Fig{fig:SQw_simulation}, we can already see that $S(\bfq,\omega)\sim 10^{-4}\, \mathrm{eV}^{-1}$ even for $q =0.1$ keV and $\omega\approx 0.01$ eV.  For reference, the rate for this value is
\beq
R\sim 1.3 \times 10^{19} \times \kappa^2 \times\left(\frac{S(\bfq,\omega)}{10^{-4} \,\mathrm{eV}^{-1}}\right)/\mathrm{kg}/\mathrm{year}.
\eeq
for a $100$ kV/cm electric field.  Even without a reliable extrapolation to $q\ll \omega$, we can therefore estimate that the rate must be very small for $\kappa$ values that satisfy current stellar constraints.

%%%%%%%%%%%%%%%%%%%%%%%
\section{Conclusions}\label{sec:Conclusions}
%%%%%%%%%%%%%%%%%%%%%%%

We have considered multi-excitation production in superfluid helium from dark matter scattering and absorption, showing that superfluid helium may be sensitive to DM in the keV to MeV mass range, with DM-nucleon cross sections between $10^{-36}$ and $10^{-44}$ $\mathrm{cm}^2$.  This extends the reach of superfluid helium beyond ordinary nuclear recoils, which can reach dark matter as light as $\sim $MeV for the same $\sim $meV energy threshold. 

 We provided an explicit calculation for the multi-excitation process, focusing on the leading two-excitation contribution to the dynamic structure function $S_m(\bfq, \omega)$. This theoretical understanding is necessary, as the existing neutron scattering data on multi-excitation production samples only a limited region in phase space for the response of the fluid. 
We calculated $S_m(\bfq, \omega)$ in a leading order approximation, which does not account for important self-interactions that modify the roton/maxon contributions and lead to substantial differences in the spectrum.  Nevertheless, we have compared this method to the re-summed numerical results in Ref.~\cite{Krotscheck2015} (CKL15), which focused on momentum transfers $q\gtrsim 100$ eV, finding that the reach for DM scattering is similar in the two cases. 
In the future, a more complete calculation of the low momentum regime, complemented with accurate measurements in neutron scattering experiments, is highly desirable.
We further calculated the rate of hidden photon absorption and hidden photon mediated dark matter scattering, both with and without an external electric field applied on the fluid. For these processes, we find that the reach is not competitive with existing stellar constraints.

Dark matter detection by multi-excitation production in superfluid helium illustrates a more general idea: by harnessing a coupling to modes other than ordinary nuclear recoils, new regimes in dark matter mass may be reached with the same technology. While we have focused on superfluid helium as a promising target, this idea warrants exploration in other types of materials.

\begin{acknowledgments}
We thank Robert Golub, Dan McKinsey, Tom Melia and Katelin Schutz for useful discussions. We are very grateful to Eckhard Krotscheck for providing us with the data of Ref.~\cite{Krotscheck2015} and for many helpful discussions on liquid helium.    The authors are supported by the DOE under contract DE-AC02-05CH11231. TL is further supported by NSF grant PHY-1316783.  This research used resources of the National Energy Research Scientific Computing Center, which is supported by the Office of Science of the U.S. Department of Energy under Contract No.DE-AC02-05CH11231.  We thank the Aspen Center for Physics, supported by the NSF Grant No. PHY-1066293, for hospitality while parts of this work were completed.
\end{acknowledgments}

%%%%%%%%%%%%%%%%%%%%%
\appendix
%%%%%%%%%%%%%%%%%%%%%

\section{Second quantization of the fluid Hamiltonian\label{app:secondquant}}

In this appendix, we provide more details on the second quantization of the fluid Hamiltonian in \Eq{eq:hydroham}, and give the details to derive \Eq{eq:H0creationbody}.  Our discussion of the formalism closely follows \cite{stephen1969raman}.

At distances longer than the inter-atomic spacing, the fluid can be described by a density field $\n(\bfr,t)$ and a (dimensionless) velocity field $\bfv(\bfr,t)$, which can be decomposed as
\begin{align}
	\n(\bfr,t) &=  \n_0 + V^{-1/2} \sum\limits_{\bfq} e^{i \bfq \cdot  \bfr} \n_{\bfq}(t), \\
	\bfv(\bfr,t)&= V^{-1/2}  \sum\limits_{\bfq}  e^{i \bfq \cdot  \bfr }  \bfv_{\bfq}(t) \,~~~
\end{align}
with $V$ the arbitrary quantization volume.
These perturbations must satisfy the continuity equation; for a classical field, the continuity equation in momentum space can be written as 
\beq\label{eq:continuity}
 \bfv_{\bfq}=\frac{i\bfq\, \dot\n_{\bfq}}{\n_0 q^2}.
\eeq
With $\mathcal{V}$ the potential energy in the fluid, the Hamiltonian of the system is
\bea\label{eq:appH}
H&=&\int d^3 \bfr\, \frac{1}{2}\mHe\, \bfv \cdot \n {\bfv}+ \mathcal{V}(\n).
\eea
By expanding in the density fluctuations, the system can be approximated as a quantum harmonic oscillator with Hamiltonian
\beq\label{eq:hamfreeapp}
H_0  &=&  \frac{1}{2}  \sum_{\bfq}   \mHe\, \n_0 \bfv_\bfq\cdot \bfv_{-\bfq} + \phi(\bfq) \n_{\bfq} \n_{-\bfq} \\
&=&  \frac{1}{2}   \sum_{\bfq} \frac{\mHe}{ \n_0 q^2} \dot\n_\bfq \dot\n_{-\bfq} + \phi(\bfq) \n_{\bfq} \n_{-\bfq}\label{eq:H0rhodot},
\eea
where $\phi(\bfq)\equiv \delta^2 \mathcal{V}/\delta\n_\bfq^2$ can be thought of as a momentum dependent force constant. The frequencies associated with the excitations are thus given by
\beq
\epsilon_0^2(\bfq) = \frac{\n_0 q^2 \phi(\bfq)}{\mHe}.
\label{eq:epsilonphi}
\eeq

This system can be quantized with the standard methods: We first compute the conjugate momentum to the density perturbation $\n_\bfq$
\beq
\pi_\bfq =\frac{\delta H_0}{\delta \dot\n_\bfq}= \frac{\mHe\dot\n_{-\bfq}}{\n_0 q^2}
\eeq
which  inserted in \Eq{eq:H0rhodot} gives
\beq\label{eq:hampi}
H_0  =  \frac{1}{2}\sum_{\bfq} \frac{\n_0 q^2}{\mHe}  \pi_{\bfq} \pi_{-\bfq} + \phi(\bfq) \n_{\bfq} \n_{-\bfq}.
\label{H0}
\eeq
We then enforce the canonical quantization condition $\left[\n_{\bfq'},\pi_\bfq\right]=i \delta_{\bfq,\bfq'}$ and decompose $\n_\bfq$ and $\pi_{\bfq}$ as 
\beq
\n_{\bfq} & = & i\sqrt{\frac{\n_0 q^2 }{2\mHe\epsilon_0(\bfq)}} \left(a_{-\bfq} - a_{ \bfq}^\dagger \right),\label{eq:rhoexpansion}\\
\pi_{-\bfq} & = & \sqrt{\frac{\mHe\epsilon_0(\bfq) }{2\n_0 q^2}} \left(a_{-\bfq} + a_{\bfq}^\dagger \right)\label{eq:piexpansion} 
\eeq 
where the $a_{\bfq}$ are the usual ladder operators, which satisfy $\left[a_{\bfq},a^\dagger_{\bfq'} \right] = \delta_{\bfq,\bfq'}$. This finally reduces the Hamiltonian to the familiar form
\beq\label{eq:harmHam}
H_0 = \sum_{\bfq} \epsilon_0(\bfq)\left(a_{\bfq}^\dagger a_{\bfq} + \frac{1}{2}\right).
\label{SHO}
\eeq
With \Eq{eq:rhoexpansion} we can also explicitly recover the Bijl-Feynman result shown in \Eq{eq:bijlfeynman}
\beq
S(\bfq)=\frac{1}{\n_0} \langle \Psi_0 | \n_{-\bfq} \n_{\bfq} |\Psi_0 \rangle=\frac{ q^2 }{2\mHe\epsilon_0(\bfq)}.
\eeq
Equivalently, we can compute
\begin{align}
 \langle\bfq|H_0 - E_0 |\bfq \rangle &= \bigg \langle \bfq \bigg| \sum_{\bfk} \left(\frac{1}{2\mHe}  \frac{\bfk^2 }{4 n_0 S(\bfk)^2} \n_{\bfk} \n_{-\bfk}+ \frac{\epsilon_0(\bfk)}{4 n_0 S(\bfk)} \n_{\bfk} \n_{-\bfk}\right) \bigg | \bfq \bigg \rangle \\
& =  \frac{1}{2} \left( \frac{\bfq^2}{2 \mHe S(\bfq)} + \epsilon_0({\bfq}) \right) =\epsilon_0(\bfq).
\label{eq:epsilonH0}
\end{align}
where we used \Eq{eq:rhoexpansion} and \Eq{eq:piexpansion} to rewrite the Hamiltonian in \Eq{H0}, while dropping terms that are annihilated by the external states.
For later usage, we also rewrite Eq.~(\ref{eq:rhoexpansion}) and (\ref{eq:piexpansion}) as
\bea
\n_{\bfq} & = & i\sqrt{\n_0 S(\bfq)} \left(a_{-\bfq} - a_{ \bfq}^\dagger \right),\label{eq:rhoexpansion2}\\
\bfv_{\bfq} & = &\frac{i\bfq}{2\mHe}\sqrt{\frac{1}{\n_0 S(\bfq)}} \left(a_{-\bfq} + a_{\bfq}^\dagger \right)\label{eq:piexpansion2} .
\eea

To compute three excitation matrix element one must include the first non-trivial term in the expansion of \Eq{eq:appH}, which is
\beq
H_1=\frac{\mHe}{2 V^{1/2}}  \sum_{\bfq,\bfk}  \bfv_{\bfq}\n_{\bfk-\bfq}\bfv_{-\bfk}
\eeq
where we neglect a possible cubic contribution from $\mathcal{V}(\n)$. In second quantized form this can be written as
\beq
H_1=-\frac{i}{8} \frac{C}{\mHe \; V^{1/2}} n_0 \sum_{q',k'l'}  \bfq'\cdot \bfl'\,S(\bfk')\, (a_{-\bfq'}+a^\dagger_{\bfq'})(a_{-\bfk'}-a^\dagger_{\bfk'})(a_{-\bfl'}+a^\dagger_{\bfl'})\delta_{\bfk'+\bfq'+\bfl',0}
\eeq
with $C\equiv1/\sqrt{n_0^3 S(\bfq)S(\bfk)S(\bfq-\bfk)}$. As discussed in \Sec{sec:freeham}, the vacuum of the free Hamiltonian in \Eq{eq:hamfreeapp} is not a good approximation of the true vacuum. The true vacuum can however be approximated to leading order in perturbation theory by
\beq\label{eq:vacshift}
|\Psi_0\rangle\approx |0\rangle +\frac{1}{H_0-E_0}H_1|0\rangle+\cdots
\eeq
where $|0\rangle$ is the vacuum of the free Hamiltonian.

The orthogonalized, excited states are defined as in Eqn.~(\ref{eqn:singleparticle}), (\ref{eqn:twoex}) and (\ref{eq:gramschmidt2}). The matrix element of interest is then
\beq
\langle\bfq|\delta H| \bfk, \bfq-\bfk\rangle=\langle\bfq|H_0-E_0| \bfk, \bfq-\bfk\rangle^0-\epsilon_0(\bfq)\langle\bfq|\bfk, \bfq-\bfk\rangle^0+\langle\bfq|H_1|\bfk, \bfq-\bfk\rangle
\eeq
where we neglected corrections of order $\mathcal{O}(H_1)$. Using \eqref{eq:vacshift}, the first term can then be written as
\bea
\langle\bfq|H_0-E_0| \bfk, \bfq-\bfk\rangle^0&=&C\Bigg[\langle 0|H_1\frac{1}{H_0-E_0}\n_{-\bfq} (H_0-E_0)\n_{\bfk}\n_{\bfq-\bfk}|0\rangle\\
&&+\langle 0| \n_{-\bfq} (H_0-E_0)\n_{\bfk}\n_{\bfq-\bfk}\frac{1}{H_0-E_0}H_1|0\rangle\Bigg]\\
&=&C\Big[r\langle 0| H_1 \n_{-\bfq} \n_{\bfk} \n_{\bfq-\bfk}|0\rangle+(1- r) \langle 0|  \n_{-\bfq} \n_{\bfk} \n_{\bfq-\bfk}H_1|0\rangle\Big]
\eea
with
\beq
r\equiv\frac{\omega_{\bfk}+\omega_{\bfk-\bfq}}{\omega_\bfq+\omega_{\bfk}+\omega_{\bfk-\bfq}}
\eeq
where the $\omega_{\bfk}$ etc are the eigenvalues of $H_0$ corresponding to the state $n_{\bfk}|0\rangle$.  Assuming that none of the external momenta are equal to one another, we find
\beq
\langle 0| H_1 \n_{-\bfq} \n_{\bfk} \n_{\bfq-\bfk}|0\rangle&=&\frac{i}{C}\langle 0| H_1 a^\dagger_{-\bfq} a^\dagger_{\bfk} a_{\bfq-\bfk}^\dagger|0\rangle\\
&=& \frac{\n_0}{4\mHe \, V^{1/2}}\left[\bfq\cdot\bfk\;S(\bfq-\bfk)+\bfq\cdot(\bfq-\bfk)\;S(\bfk)-\bfk\cdot(\bfq-\bfk)\;S(\bfq)\right] \nonumber
\eeq
We moreover have 
\bea
\langle 0|  \n_{-\bfq} \n_{\bfk} \n_{\bfq-\bfk}H_1|0\rangle&=&\langle 0|  \n_{-\bfq} \n_{\bfk} \n_{\bfq-\bfk}H_1 |0\rangle^\dagger\\
&=&\langle 0| H_1\n_{\bfk-\bfq}\n_{-\bfk}\n_{\bfq}  |0\rangle\\
&=&\langle 0| H_1 \n_{-\bfq} \n_{\bfk} \n_{\bfq-\bfk}|0\rangle
\eea
where in the last line we used that all $n$ commute and that the theory is parity invariant. Putting all of this together, we then have
\bea
\langle\bfq|(H_0-E_0)| \bfk, \bfq-\bfk\rangle^0=\frac{\n_0 C}{4\mHe \, V^{1/2}}\Big[\bfq\cdot\bfk\;S(\bfq-\bfk)+\bfq\cdot(\bfq-\bfk)\;S(\bfk)-\bfk\cdot(\bfq-\bfk)\;S(\bfq)\Big]. \nonumber
\eea
Similarly, we can compute $\langle\bfq|H_1|\bfk, \bfq-\bfk\rangle$. In this case the relevant term in the Hamiltonian is
\beq
H_1=-\frac{i n_0 C}{8\mHe\, V^{1/2}} \sum_{q',k',l'}\Big[ 2\bfq'\cdot\bfl'\;S(\bfk')-\bfk'\cdot\bfl' S(\bfq')\Big]a^\dagger_{\bfq'} a_{-\bfk'} a_{-\bfl'}\; \delta_{\bfq'+\bfk'+\bfl',0} + \cdots
\eeq
To leading order in $H_1$, this matrix element is then
\beq
\langle\bfq|H_1|\bfk, \bfq-\bfk\rangle&=&C\langle 0|  \n_{-\bfq} H_1 \n_{\bfk} \n_{\bfq-\bfk}|0\rangle\\
&=&-i\langle 0|  a_{\bfq} H_1 a^\dagger_{\bfk} a_{\bfq-\bfk}^\dagger|0\rangle\\
&=& \frac{\n_0 C}{4\mHe V^{1/2}}\left[\bfq\cdot\bfk\;S(\bfq-\bfk)+\bfq\cdot(\bfq-\bfk)\;S(\bfk)+\bfk\cdot(\bfq-\bfk)\;S(\bfq)\right].
\eeq
The final result is
\beq
\langle\bfq|(H_0+H_1-E_0)| \bfk, \bfq-\bfk\rangle^0=\frac{\bfq\cdot\bfk\;S(\bfq-\bfk)+\bfq\cdot(\bfq-\bfk)\;S(\bfk)}{2\mHe\sqrt{N S(\bfk)S(\bfq)S(\bfq-\bfk)}}.
\eeq
which matches the result in \Eq{eq:matrixelstep1}, which was performed in the microscopic formalism. 

One may also attempt to compute the overlap term $\langle\bfq|\bfk, \bfq-\bfk\rangle^0$ is this quantum fluid formalism. This however gives an answer which differs from the convolution approximation, as computed in \App{app:overlap}. This is unsurprising, since the fluid Hamiltonian is merely a low energy effective theory, which in itself does not capture the full UV dynamics. Ideally, one would address this by computing the relevant matching terms from directly coarse-graining the microscopic physics. In the absence of such microscopic understanding, we estimate the overlap term with the heuristic ansatz provided by the convolution approximation. 

\section{Derivation of the overlap term \label{app:overlap}}

In this appendix we provide a derivation for the overlap term in \Eq{eq:overlapterm}. As discussed in \Sec{sec:threephononvertex}, the $^0\langle\bfk-\bfq,\bfk|\bfq\rangle$ overlap term encodes aspects of the dynamics of the strongly coupled fluid, and is not known from first principles. It is however possible to constrain it with a number of consistency conditions and subsequently derive a closed form expression by choosing an ansatz for the remaining unknown part. The ansatz we work with here is known as the convolution approximation, and our derivation closely follows the discussion in Ref.~\cite{JacksonFeenberg}.

In integral form, the overlap term can be written as
\bea\label{eq:overlapintapp}
	\leftidx{^0}\langle \bfq - \bfk, \bfk | \bfq \rangle = \frac{1}{\sqrt{n_0^3 S(\bfq-\bfk) S(\bfq) S(\bfk)} }\int d^3\bfr_1 ... d^3 \bfr_N\, \psi_0 \n^*_{\bfq-\bfk} \n^*_{\bfk}   \n_{\bfq} \psi_0.
\eea
Using \Eq{eq:densitydef}, we can rewrite the density operators as follows
\begin{align}
\n^*_{\bfq-\bfk} \n^*_{\bfk}   \n_{\bfq}=& V^{-3/2}\sum_{m,n,p} e^{i \bfk\cdot(\bfr_m-\bfr_n)+i \bfq\cdot(\bfr_p-\bfr_m)}\\
=&V^{-3/2}\Bigg[N+ \sum_{\substack{m,n\\
                  m\neq n}} e^{i \bfk\cdot(\bfr_m-\bfr_n)}+ \sum_{\substack{m,p\\
                  m\neq p}} e^{i \bfq\cdot(\bfr_p-\bfr_m)}\\
&+ \sum_{\substack{p,n\\
                  p\neq n}} e^{i (\bfq-\bfk)\cdot(\bfr_p-\bfr_n)}+ \sum_{\substack{m,n,p\\m\neq n\neq p }} e^{i \bfk\cdot(\bfr_m-\bfr_n)+i \bfq\cdot(\bfr_p-\bfr_m)} \Bigg]\\
\rightarrow V^{-3/2}&\Bigg[-2N+ \big|\sum_n e^{i \bfk\cdot \bfr_n}\big|^2+\big|\sum_n e^{i \bfq\cdot \bfr_n}\big|^2 + \big|\sum_n e^{i (\bfq-\bfk)\cdot \bfr_n}\big|^2+ N^3 e^{i \bfq\cdot\bfr_{12}+i \bfk\cdot\bfr_{23}} \Bigg]
\end{align}
with $\bfr_{ij}\equiv\bfr_i-\bfr_j$. In the last term we collected equivalent terms under the integral. We hereby used that $\psi_0$ is assumed to be invariant under permutations of the $\bfr_i$ and we took $N\approx N-1\approx N-2$. If substituted in \Eq{eq:overlapintapp}, this results in
\begin{align}
\leftidx{^0}\langle \bfq - \bfk, \bfk | \bfq \rangle =&\frac{1}{\sqrt{N S(\bfq-\bfk) S(\bfq) S(\bfk)} }
\Bigg[-2 +S(\bfq)+S(\bfk)+S(\bfq-\bfk)\nonumber\\
&+ \frac{1}{N}\int d^3\bfr_1d^3\bfr_2d^3\bfr_3\, e^{i \bfq\cdot\bfr_{12}+i \bfk\cdot\bfr_{23}}\, p_3( \bfr_1, \bfr_2,\bfr_3)\Bigg]\label{eq:overlapintermediate}
\end{align}
with 
\beq
p_3(\bfr_1,\bfr_2,\bfr_3)\equiv N(N-1)(N-2)\int\! d^3\bfr_4... d^3\bfr_N\, \psi_0^2.
\eeq
The function $p(\bfr_1,\bfr_2,\bfr_3)$ is usually referred to as the three-particle distribution function. Similarly, we can define the two-particle distribution function
\beq
p_2(\bfr_1,\bfr_2)\equiv N(N-1)\int\! d^3\bfr_3... d^3\bfr_N\, \psi_0^2.
\eeq
The Fourier transform of the two-particle distribution function is closely related to the static structure function, in particular
\bea
S(\bfq)&=&\frac{1}{\n_0}\int\!\! \psi_0^2\, \n^\ast_{\bfq}\,\n_{\bfq}\, d^3 \bfr_1...d^3\bfr_N\\
&=&1+\frac{1}{N}\int\!\! \, p_2(\bfr_1,\bfr_2)\,e^{i\bfq\cdot \bfr_{12}} d^3\bfr_1d^3\bfr_2\label{eq:Stwopart}
\eea
where the first term comes from the terms in the sum with $k=l$.

Assuming translation invariance, the two and three particle distribution functions must satisfy the following recursion relations
\bea
\frac{1}{N-1}\int d^3\bfr_{2}\, p_2(\bfr_{1},\bfr_2)=p_1(\bfr_1)=\n_0\\
\frac{1}{N-2}\int d^3\bfr_{3}\, p_3(\bfr_{1},\bfr_2,\bfr_3)      =p_2(\bfr_{1},\bfr_2).\label{eq:threepointrecursion}
\eea
In particular \Eq{eq:threepointrecursion} allows us to constrain the three-particle distribution function. It is convenient to define the dimensionless function 
\beq
h(\bfr_{12})=\frac{1}{\n_0^2}p_2(\bfr_1,\bfr_2)-1.
\eeq
Without loss of generality, we can decompose the three particle distribution function as
\bea\label{eq:threepointsfactor}
p_3(\bfr_{1},\bfr_2,\bfr_3)&=&\n_0^3\Big[ 1+h(\bfr_{12})+h(\bfr_{23})+h(\bfr_{13})+h(\bfr_{12})h(\bfr_{23})\nonumber\\
&&+h(\bfr_{23})h(\bfr_{31})+h(\bfr_{31})h(\bfr_{12}) \Big]+\delta p_3(\bfr_{1},\bfr_2,\bfr_3).
\eea
The term in the brackets models the behavior of $p_3(\bfr_{1},\bfr_2,\bfr_3)$ when two or more points are well separated, and the non-factorized core $\delta p_3(\bfr_{1},\bfr_2,\bfr_3)$ captures the UV behavior and is large when all three points are close together. To satisfy \Eq{eq:threepointrecursion} one must require
\beq\label{eq:corecondition}
\int \delta p_3(\bfr_{1},\bfr_2,\bfr_3) d^3\bfr_3=-\n_0^3\int h(\bfr_{13})h(\bfr_{23}) d^3\bfr_3.
\eeq

At this point in the derivation it becomes necessary to choose an ansatz for $\delta p_3(\bfr_{1},\bfr_2,\bfr_3)$. A popular choice is the \emph{convolution approximation}, where we take 
\beq \label{eq:coreansatz}
\delta p_3(\bfr_{1},\bfr_2,\bfr_3)=\n_0^4 \int  h(\bfr_{14})h(\bfr_{24}) h(\bfr_{34})d^3\bfr_4
\eeq
which satisfies \Eq{eq:corecondition}. Substituting \Eq{eq:threepointsfactor} in the integral in \Eq{eq:overlapintermediate}, we find that the first four terms are of the form
\bea
\int d^3\bfr_1d^3\bfr_2d^3\bfr_3\, e^{i \bfq\cdot\bfr_{12}+i \bfk\cdot\bfr_{23}}\times 1 &=&V\, \delta^{(3)}(\bfk)\,\delta^{(3)}(\bfq)\label{eq:vanish1}\\
\int d^3\bfr_1d^3\bfr_2d^3\bfr_3\, e^{i \bfq\cdot\bfr_{12}+i \bfk\cdot\bfr_{23}} h(\bfr_{12}) &=&\frac{N}{\n_0^2}\delta^{(3)}(\bfk)\big(S(\bfq)-1\big)-V\, \delta^{(3)}(\bfk)\,\delta^{(3)}(\bfq)\label{eq:vanish2}
\eea
plus permutations. In the second line we used \Eq{eq:Stwopart}. These terms all vanish, since we are interested in $\bfk,\bfq\neq0$. The next three terms are of the form
\bea
\int d^3\bfr_1d^3\bfr_2d^3\bfr_3\, e^{i \bfq\cdot\bfr_{12}+i \bfk\cdot\bfr_{23}} h(\bfr_{12}) h(\bfr_{23}) &=&\int d^3\bfr_1d^3\bfr_2d^3\bfr_3\, e^{i \bfq\cdot\bfr_{12}+i \bfk\cdot\bfr_{23}} \frac{p_2(\bfr_1,\bfr_2)}{\n_0^2} \frac{p_2(\bfr_2,\bfr_3)}{\n_0^2} \nonumber \\
&=&\frac{N}{\n^3_0} \big(S(\bfk)-1\big)\big(S(\bfq)-1\big)
\eea
where we used \Eq{eq:Stwopart} and dropped terms of the form in Eqs.~(\ref{eq:vanish1}) and (\ref{eq:vanish2}). Similarly, inserting \Eq{eq:coreansatz} results in
\begin{equation}\hspace{-.2cm}\resizebox{ \textwidth}{!} 
{ $\int d^3\bfr_1...d^3\bfr_4\, e^{i \bfq\cdot\bfr_{12}+i \bfk\cdot\bfr_{23}} h(\bfr_{14})h(\bfr_{24}) h(\bfr_{34}) = \frac{N}{\n_0^4} \big(S(\bfk)-1\big)\big(S(\bfq)-1\big)\big(S(\bfk-\bfq)-1\big).$
}
\end{equation}
Putting everything together then finally gives
\beq
\langle \bfq-\bfk, \bfk | \bfq \rangle ^0 = \frac{\sqrt{ S(\bfq-\bfk) S(\bfq) S(\bfk)}}{\sqrt{N}}.
\eeq

\bibliographystyle{jhep}
\bibliography{HeDM}

\end{document}